%% file: paper.tex
\documentclass[]{article}

\pdfoutput=1
\usepackage{setspace} 

\catcode`\@=11\relax
\usepackage{fancyhdr,timestamp}
\usepackage{amsthm,amssymb}
\usepackage{latexsym}

\usepackage{algorithm,algorithmic}

\usepackage{rotating} 

\input{definitions}

\usepackage{graphicx}
\graphicspath{{figs/}}

\title{ {\bf Distance Maps and Plant Development \#2: Facilitated Transport
    and Uniform Gradient} } 

\author{Pavel Dimitrov and Steven W. Zucker} 

\date{\today}

\begin{document}

\maketitle

\begin{abstract}
The principles underlying plant development are extended to allow a
more molecular mechanism to elaborate the schema by which ground cells
differentiate into vascular cells.  Biophysical considerations dictate
that linear dynamics are not sufficent to capture facilitated auxin
transport (e.g., through PIN). We group these transport facilitators
into a non-linear model under the assumption that they attempt to
minimize certain {\em differences} of auxin concentration.  This
Constant Gradient Hypothesis greatly increases the descriptive power
of our model to include complex dynamical behaviour. Specifically, we
show how the early pattern of PIN1 expression appears in the embryo,
how the leaf primordium emerges, how convergence points arise on the
leaf margin, how the first loop is formed, and how the intricate
pattern of PIN shifts during the early establishment of vein patterns
in incipient leaves of Arabidopsis. Given our results, we submit that
the model provides evidence that many of the salient structural
characteristics that have been described at various stages of plant
development can arise from the uniform application of a small number
of abstract principles.
\end{abstract}

\tableofcontents

\newpage

\section{Introduction}




In the companion paper \cite{plos-paper1} we introduced a model for
auxin dynamics that can predict key features of leaf vein
patterns. While it is remarkable that diffusion of the hormone
provides a sufficient basis to describe these patterns, this transport
mechanism does not accurately describe the flow of auxin in
plants. Facilitated transport of the hormone has been postulated since
the proposal of the chemiosmotic theory, and recent molecular genetic
and cell biological findings support this theory. Genes and related
proteins have recently been identified as either influx (e.g., AUX) or
efflux (e.g., PIN) auxin `carriers'. Most interestingly, the
experiments suggest that the hormone may in fact be `pushed' against a
concentration gradient --- that is, opposite to the direction of
diffusion flow in the sense of our earlier model. The carriers also
appear to increase flow in the direction of diffusion flow at times,
but why this dual function exists and when it shifts from one to the
other is not well understood. Opinion among experimentalists is split,
resulting in a conundrum for plant developmental biology.  Recent
experiments that focus on facilitated transport due to the PIN1 gene
provide another dimension to the
conundrum. Scarpella~\etal~\cite{scarpella:genes-for-procambium}
visualize PIN1 expression domain, or PED, at different stages of early
leaf development and observe that eventually the PED defines the vein
pattern exactly. Curiously, the PED undergoes several transformations
whereby certain parts appear temporarily only to disappear just before
the final pattern is established. Yet, the patterns that the PED
defines are robust to cell divisions. The dynamics are non-linear and
sometimes counter-intuitive. It would appear, therefore, that a
carefully timed complex genetic machinery must exist for the purpose
of controlling these events.

Contrary to this intuition, in this paper we demonstrate that the
underlying principles that guide such intricate events need not be
many. While our focus remains abstract, our goal is to bring our model
closer to cell and molecular biology. We concentrate on the phenomenon
of polar transport rather than its molecular implementation and extend
our earlier model to accommodate this abstraction. Building on the two
principles from our earlier work, the Constant Production Hypothesis
and the Proportional Destruction Hypothesis, we develop the Constant
Gradient Hypothesis, which we use to derive a new computational tool
to test the theoretical predictions. We introduce a non-linear model
mathematically and then use it to analyze the observations reported by
Scarpella~\etal~\cite{scarpella:genes-for-procambium}. We focus on the
shifting patterns of the PED during early leaf development, and
discuss schematic simulations to offer an explanation for key stages
of those events. Thus we illustrate how our theory captures certain
intricate patterning events. Insofar as the polar transport mechanism
is implemented by a single substance, our simulations predict the
behavior of that substance; however, (and this is key) our polar
transport extension is meant as a simplifying abstraction of the
Chemiosmotic Theory that captures the total effect of various
substances acting together, rather than the behavior of any one
substance. To preview the scope of our predictions, we sketch in
\reffig{fig:fi1-sketch} certain of the pivotal events in vascular
development. Once our theory is developed, we return to these
predictions via simulation.

\begin{figure}

\caption{Sketch of predictions.}
\label{fig:fi1-sketch}
\end{figure}

\section{Auxin Transport}


Auxin travels through plants by moving between cells. It may leave a
cell interior, the cytoplasm, pass through the plasmalemma and then
reach the cell walls, or apoplast. It can then move into the cytoplasm
of an adjacent cell and continue diffusing in this fashion throughout
the plant. Perhaps the first real breakthrough in understanding how
this movement occurs was the introduction of the chemiosmotic theory
in the 1970s. Based only on physical chemistry and a handful of
measurements, Rubery and
Sheldrake~\cite{rubery-sheldrake-1973,rubery-sheldrake-1974} and
Raven~\cite{raven-1975} were able to predict that auxin cannot travel
by diffusion alone; that some sort of active transport (i.e. one that
requires energy expenditure) must be present; and that some sort of
carrier substances must exist that move auxin in a preferred
direction---toward the cytoplasm (influx) or away from it (efflux).

\begin{table}

  \begin{tabular}{r|c|c|l}

    Protein & Polar Membrane Localization &  Auxin
    ``carrier'' & References\\

    PIN1 & Yes & Efflux & \cite{okada-etal-1991},
    \cite{galwiler-etal-1998} \\

    PIN2 
    & Yes & Efflux & \cite{chen-etal-1998},
    \cite{luschnig-etal-1998}, \cite{muller-etal-1998},
    \cite{utsuno-etal-1998}, \cite{sieberer-etal-2000},
    \cite{abas-etal-2006}, \cite{blilou-etal-2005} \\

    PIN3 & -- & Efflux & \cite{friml-etal-2002a} \\

    PIN4 & Yes & Efflux & \cite{friml-etal-2002b},
    \cite{sabatini:auxin-in-roots} \\

    PIN7 & -- & Efflux & \cite{friml-etal-2003} \\

    MDR1/ PGP\{1,2,4,19\} & Yes and No & Efflux &
    \cite{brown-etal-2001}, \cite{gil-etal-2001},
    \cite{murphy-etal-2000}, \cite{murphy-etal-2002},
    \cite{noh-etal-2001}, \cite{noh-etal-2003} \\

    AUX1 & Yes & Influx & \cite{bennet-etal-1996},
      \cite{swarup-etal-2001}, \cite{marchant-etal-2002}
  \end{tabular}

  \caption{\label{tab:auxin-carriers} The variety of auxin
    facilitators reveals enormous potential for complex
    interactions. Since many of these remain to be studied, we study
    their net effect.}
\end{table}

It is ``remarkable how accurately [this] molecular model [...]  fits
with the recent molecular genetic and cell biological findings''
\cite{vieten-etal-2007}. Increasingly, the evidence suggests that
auxin transport is facilitated in both directions. The AUX proteins
are a family of putative influx carriers
\cite{marchant-etal-2002,swarup-etal-2004,dharmasiri-etal-2006,yang-etal-2006,kerr-bennet-2007},
while the PIN family \cite{kerr-bennet-2007,vieten-etal-2007} and the
multidrug resistance/p-glyvoprotein (MDR/PGP) family of proteins
\cite{blakeslee-etal-2005} are thought to be efflux carriers. Recent
molecular techniques, which can effectively `color' specific molecules
with sub-cellular precision \cite{friml-palme-2002}, have shown that
both types of facilitators may localize asymmetrically on the
plasmalemma although MDR/PGP proteins do not always do that
\cite{santelia-etal-2005}. Indeed, many proteins have been implicated
in auxin transport, as Table~\ref{tab:auxin-carriers} summarizes, and
often more than one substance is present in the same cell at the same
time. It is therefore difficult to infer the pattern of auxin
transport from this detailed mechanistic view of auxin transport.

In order to analyze how the patterns of auxin transport form we shall
consider polar transport in a cell as the total effect of polar
transport as predicted by the chemiosmotic theory. So, for example, if
both AUX proteins and PIN proteins are expressed in a cell in equal
measure and on the same parts of the membrane, then the action of the
influx carriers will be canceled by the action of the efflux carriers
and the total effect will be nil. Conversely, if we observe a total
efflux effect in a given cell, then this only implies that the balance
between influx and efflux carriers favors the latter. This
simplification requires a refinement  to our earlier model, which
we prove to be consistent with the Chemiosmotic theory in
Appendix~\ref{sec:chemiosmotic-active-transport}. In particular, we
may limit the analysis to the movement of auxin between cell interiors
and yet capture a sufficiently accurate picture of the whole
process. We note that, if the molecular carriers in a given region of
cell are always of the same kind, as seems to be the case in
Scarpella~\etal~\cite{scarpella:genes-for-procambium} (all PIN1), then
any predictions using this simplification apply directly to that
carrier. For this reason and because most of the experiments that we
shall analyze involve only PIN proteins, we shall refer to our
simplification of directed (polar) transport as ``PIN'' transport.

The key difference between the model that we develop here and our
earlier formulation is that polar transport may move auxin against
concentration gradients (from low to high concentrations) and
therefore counteract the passive diffusion flow. The modeling
challenge thus moves beyond that of Fickian diffusion, but its
importance increases dramatically. The reward is that the capability
for pattern formation is enlarged as well. We now explore this
increased capability.




\section{``PIN'' Polarity}
\label{sec:pin-polarity}

We now consider some of the experimental evidence concerning
PIN1. This will be the basis for our assumptions about polar transport
recognizing the fact that the experiments reviewed below show PIN1 but
other carriers may be present in the same cells.


\subsection{A Conundrum}


The appearance of PIN1 in detectable quantities is intimately related
to auxin \cite{vieten-etal-2005,paciorek-etal-2005}, but the polarity
of the carrier need not follow the diffusive auxin flow. For example,
Reinhardt~\etal~\cite{reinhardt-etal-2003} report that the pattern of
PIN expression changes after a micro-application of auxin in shoot
meristems---the polar transport acts toward the application site. In
leaves, on the other hand, Scarpella~\etal~\cite{berleth:bipolarcell}
applied the hormone to a small portion of a developing primordium and
noticed that PIN1 oriented so that the hormone would be better drained
away from the leaf, i.e. away from the application site. The
application site in both cases has a higher auxin concentration than
the rest of the organ so, considering each experiment separately, two
mechanisms come to mind:
\begin{description}
  \item[{\sc Mechanism 1:}] New PIN1 appears at a membrane against
    diffusion flow when this flow, or difference of concentration, is
    large enough.

  \item[{\sc Mechanism 2:}] New PIN1 appears at a membrane `helping'
    diffusion flow.
\end{description}

At first glance it looks like neither rule can explain both
experiments because the phenomena appear to be mutually exclusive. But
it turns out that one of these schematic mechanisms is the foundation
of a model that predicts both phenomena in detail as well as vein
patterns and the growth of organs more generally.

\begin{figure}[t]

\begin{center}
  \begin{tabular}{cc}
    \includegraphics{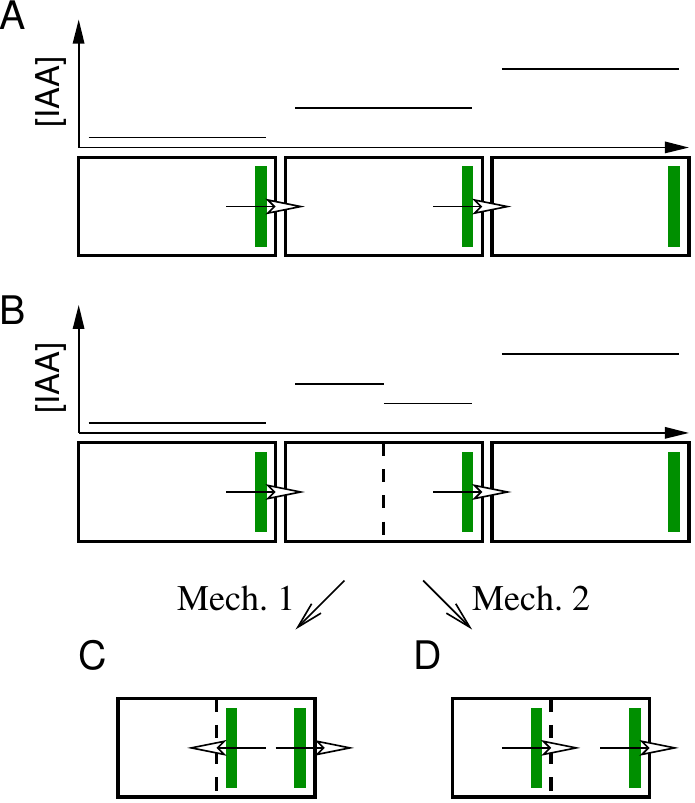} &
    \includegraphics{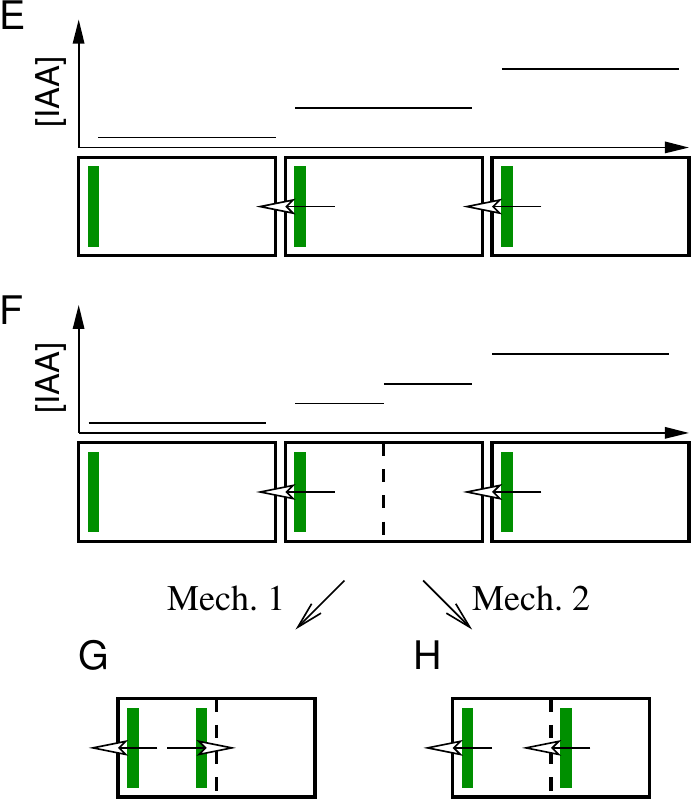}
  \end{tabular}
\end{center}

\caption[Illustration of the effect of two presumptive mechanisms for
  PIN1 creation in the context of a cell
  division.]{\label{fig:pin-polarity} Illustration of the effect of
  two presumptive mechanisms for PIN1 creation in the context of a
  cell division. Three cells are shown in \CP{A} and \CP{E} with
  consistent polarities indicated by open arrows. When the middle cell
  divides, its membrane remains the same except for the new membrane
  that splits the cell in two daughter compartments. We take this to
  imply that the existing PIN structure is maintained and the problem
  is to determine how PIN is established along the new section of
  membrane. The concentration of auxin (denoted [IAA]) is plotted
  above the depiction of each cell. PIN1, shown in green, is created
  according to one of two mechanisms: (1) toward higher concentrations
  (see \CP{C} and \CP{G}), or (2) toward lower concentrations (see
  \CP{D} and \CP{H}). Note that Mechanism 1 breaks the continuity of
  PIN polarity in both cases, whereas Mechanism 2 maintains this
  continuity. \FP{A--D} PIN1 acts against diffusive flow, toward
  increasing concentration, e.g. as in the {\em Arabidopsis} root
  tip. \FP{E--H} PIN1 `helps' diffusive flow, e.g. as in leaf
  primordia. \FP{A,E} Configuration before cell division. \FP{B,F}
  Configuration during cell division (dashed line) but before PIN1 is
  established at the new interface. \FP{C,D,G,H} Configuration in
  middle cell (and respective daughter cells) after PIN1 is created
  according to Mechanism~1 (\CP{C,G}) or Mechanism~2 (\CP{D,H}).}
\end{figure}

\subsection{PIN1 Polarity Under Cell Division}
\label{sec:pin-polarity-celldiv}

Our first clue comes from considering what happens when a cell with
PIN1 expression divides. In leaf primordia, the cells that will
eventually become part of the midvein are marked by PIN1 expression
that localizes toward the base of the cell, i.e. toward the stem and
away from the leaf tip. This pattern appears as soon as the primordium
emerges from the shoot apical meristem and the polarity of the cells
is maintained even during frequent cell divisions. When a cell with
PIN1 expression divides perpendicularly to the midvein strand, a new
membrane is created that also acquires PIN1 expression with the
polarity of the original cell. Mechanism~1 fails to predict such
behavior, while Mechanism~2 maintains the proper PIN1 polarity. We
reason as follows.

Consider the illustration in
\reffig[A--D]{fig:pin-polarity}. Initially, the concentration of auxin
in the middle cells is higher than in the cell on the left and lower
than in the cell on the right. When the middle cell divides its
membrane is `split' and a new barrier forms. The existing auxin
carriers then drain auxin away from the daughter compartment on the
right and push auxin into the neighbor cell on the left. On the other
hand, the carriers from within the neighbor cell on the left push
auxin into the daughter compartment on the left. Thus, the left
compartment acquires a higher concentration than the right
compartment. If PIN1 were to be created toward a higher
concentration---Mechanism~1 and \reffig[C]{fig:pin-polarity}---then it
should form in the right compartment thereby creating a bipolar
cell. This would be the outcome after most cell divisions and, in
particular, in the midvein strand where no bipolar cells have been
found. By contrast, if PIN1 were to be created predominantly toward
lower concentrations of auxin---Mechanism~2 and
\reffig[D]{fig:pin-polarity}---then the continuity would be
maintained. We therefore conclude that the first appearance of PIN1 in
cells is most likely due to Mechanism~2. 

Recall Schema 1 from \cite{plos-paper1} which postulates that a ground
cell becomes c-vascular whenever it measures a sufficiently large
$\Delta c$. At the schema level of abstraction, the change is realized
as an increased diffusion coefficient through the relevant interface,
and the $\Delta c$ there decreases. This is precisely the effect of
Mechanism~2!  In other words, we have now shown that Schema 1 can be
implemented by auxin carriers. Therefore our derivation of Schema 1
informs the appearance of auxin carriers as a result of organ geometry
and cell size distribution. The agreement between these predictions
and experimental fact reinforces confidence that Schema 1 is
implemented by these carriers.

\section{The Constant Gradient Hypothesis}

Even though Mechanism~2 will be a useful guideline to understand PIN1
dynamics, it does not explain how the carrier concentration at the
cell membrane is maintained. Indeed, the protein cycles between the
interior of the cell and the plasmalemma \cite{paciorek-etal-2005},
and it is somehow destroyed during the early stages of leaf
development \cite{berleth:bipolarcell} and when it undergoes
reorientation \cite{reinhardt-etal-2003}. Moreover, even if the first
appearance of PIN1 is in the direction of decreasing auxin
concentration, there is carrier presence against the auxin
gradient. For example, the polarity of provascular cells near the
distal tip of the {\em Arabidopsis} root is toward the tip, and the
auxin concentration increases in the same direction
\cite{kerr-bennet-2007}. This suggests that the mechanism responsible
for PIN1 density at the plasmalemma exhibits some sort of hysteresis
memory. Taken together, the cycling, the plasticity of expression, and
the memory properties of PIN1 imply that the mechanism is a dynamical
system governed by a differential equation.

Mathematical formulations of this sort are common and find many
applications in biology \cite{ptashne-2004}. For example, they are the
theoretical basis for genetic switches \cite{alon-2007} where the
equations can be understood in terms of two competing processes: the
activity of promoters described by a production function, and the
activity of inhibitors and natural degradation described by a
destruction function. The production function depends on the
concentration of the gene product whenever the product promotes the
expression of the gene, and it can usually be described by a Hill
function. If there are several promoter sites and cooperation between
them, then the Hill coefficient is large
\cite{goldbeter-koshland-1981, garcia-etal-2007,
  monod-etal-1965,eigen-1967}. On the other hand, the destruction is a
linear function of concentration whenever no inhibitors exist and only
natural degradation depletes the substance.  Configurations of this
kind exhibit the behavior of a switch: if an external event lowers the
concentration below a threshold $\tau$, then the system maintains the
concentration at level $A$; if another external event pushes the
concentration above this threshold then the system maintains the
concentration at level $B$ ($B>\tau>A$). If $A$ is biologically
negligible (insufficient for additional reactions) but $B$ is not,
then we can say that the gene is turned `off' or `on' depending on
whether the product concentration is below or above the threshold,
respectively.

There is more direct and independent support for the dynamics of PIN1
to be well described by a Hill function and by the concept of
cooperation. Cooperation is central in the empirically derived
Canalization Hypothesis by Sachs~\cite{sachs-1981} whereby ``flux
begets flux.'' If the increase in flux is the result of ``PIN'', then
the carriers must exhibit an autocatalytic behavior because they are
saturable: they have a maximum capacity for transporting auxin through
the plasmalemma \cite{petrasek-etal-2006,vieten-etal-2007}. Note,
however, that we are referring to the net effect and that the actual
behavior may involve more than a single substance. For example, it may
be the result of mutual cooperation between the PIN proteins and the
MDR/PGP proteins \cite{noh-etal-2003,petrasek-etal-2006}.  Therefore
to increase the flux of auxin due to those facilitators, the
concentration of the molecule must be increased. We submit the
following hypothesis:



\begin{hypothesis}[PIN1 Cooperation]
  \label{hypothesis:pin-dynamics}
  PIN carriers at the plasmalemma attract additional PIN carriers near
  them. The protein leaves the membrane when it has transported its
  load of auxin.
\end{hypothesis}

\noindent
In other words, the density of PIN carriers at a patch of the cell
plasma membrane is determined by a dynamical system with a production
function that depends on that density, a Hill function. The
destruction function also depends on the density but can be assumed to
be linear, because only a portion of the loaded proteins---those
immediately adjacent to the membrane, those attached to it---can
facilitate hormone transport at any given time. In effect, this
configuration constitutes an auxin carrier `switch.'

As we argued above, the external events that turn the switch `on' or
`off' depend on auxin and, particularly, on the difference of auxin
concentration $\Delta c$ through an interface. The rule which we
called Mechanism~2 above implies that a large $\Delta c$ in the
direction of diffusion---i.e. a positive $\Delta c$---should increase
the production function. On the other hand, the reorientation of the
carrier following external application of auxin implies that a high
gradient against the action of PIN should increase the destruction
function. Yet, the strong expression of PIN1 against the auxin
gradient in provascular cells of the root tip implies that the
destruction function overwhelms the production function only when
$\Delta c$ is sufficiently negative. We therefore submit that PIN
dynamics are modified by auxin in the following fashion:\footnote{In
  fact, a more mechanistic hypothesis that justifies
  Hypothesis~\ref{hypothesis:dc-effect-on-pin-dynamics} can be
  made. It is about both influx (e.g., AUX) and efflux (e.g., PIN)
  carriers and requires them to function in a dual fashion: $\Delta c$
  makes it difficult for PIN to unload auxin and easy for AUX. Here,
  $\Delta c$ is the more immediately available difference of auxin
  concentration between the cytoplasm and the cell wall instead of, as
  we conjecture in
  Hypothesis~\ref{hypothesis:dc-effect-on-pin-dynamics}, a $\Delta c$
  between the cytoplasms of two neighboring cells.}

\begin{hypothesis}[Effect of Auxin on PIN1 Dynamics]
  \label{hypothesis:dc-effect-on-pin-dynamics}
  PIN accumulation on the plasmalemma is promoted by $\Delta c$
  regardless of sign. The destruction function increases only when the
  diffusive flow opposite to the action of PIN1 increases, i.e. if the
  carrier pushes auxin against diffusion.
\end{hypothesis}




In other words, PIN1 is more likely to be destroyed when it pushes
auxin against a concentration gradient. But even in that case, the
production process may still be sufficiently strong to compensate for
the action of the destruction process and maintain carrier presence
or, as in the root tip, even increase it. However, if the gradient is
too steep, then destruction overwhelms production and carrier presence
disappears. 

The lack of experimental data prevents us from determining exactly
when this happens and the explicit forms of the functions that
describe the dynamics. This underlines the advantage of our
abstraction, though: we can use our theory to infer a functional role
for auxin carriers that is consistent with our hypotheses and that
will allow us to predict the distribution of polar transport intensity
in plants. Some data of this sort are available, and more precise
measurements can be obtained with current techniques.

Recall that we required veins to possess an efficient transport
mechanism for auxin. In particular, no concentration peaks should form
there. On the other hand, we saw that the appearance of auxin carriers
at a cell interface is most likely triggered by a high $\Delta c$ and,
initially at least, helps diffusion. In effect, carriers decrease this
$\Delta c$. Working against diffusion, too, they maintain a $\Delta c$
that does not exceed some fixed threshold because otherwise carrier
presence would disappear. Therefore, we propose a third uniformity
principle:
\begin{hypothesis}[Constant Gradient]
  \label{hypothesis:cgh}
  Facilitated transport maintains the difference in auxin
  concentration, $\Delta c$, constant.
\end{hypothesis}

\section{Computational Model of Active Transport}
\label{sec:non-linear-model}

The Constant Gradient Hypothesis suggests an elaboration of the
computational model developed in the previous paper
\cite{plos-paper1}. In this section we derive the mathematical and
computational tools that are necessary to evaluate the theory. We
develop a framework for simulations, which we use in the next section
to illustrate how to explain experimental evidence of certain
patterning phenomena. The tools that we develop here have a larger
scope than is shown in the next section, as can be seen in the next
paper \cite{plos-paper3} of this series.

\subsection{Background}
Even though auxin transport may be due to energy expenditure,
the net effect may still be well described by Fick's law. This is the
case when carriers are located in a non-polar fashion (homogeneously)
along the cell membrane. The PGP family appears to exhibit this
behavior so it can be seen as a mechanism for improving the effective
diffusion coefficient. In this case, our Helmholtz formulation from
\cite{plos-paper1} applies and large domains of cells can be
described by the continuous equation
\begin{equation}
  \frac{\del c}{\del t} = D \grad^2 c + \frac{K}{S} - \alpha c ~.
\end{equation}

\noindent
Thinking of cells as individual compartments, this means that the
concentration $c(i)$ of auxin in each cell $i$ changes as a function
of time according to the transport to neighboring cells $Nbr(i)$, the
production $K/S(i)$, and the destruction $-\alpha c(i)$; in symbols:
\begin{equation}
  \label{eq:discrete-helmholtz}
  \frac{\del c(i)}{\del t} = \underbrace{\sum_{j\in Nbr(i)} D
    (c(j)-c(i))}_\mathrm{Diffusion~Transport} + \frac{K}{S(i)} -
  \alpha c(i) ~.
\end{equation}

\noindent

On the other hand, the PIN family of proteins (and PIN1 in particular)
localize asymmetrically along the membrane---with negligible or no
expression on one side of the cell and strong expression on the
other. This is the scenario that we now study
because it yields dynamics for auxin transport that deviates from
Fick's Law: circumstances exist in which it can `push' the hormone
against concentration gradients. A typical assumption regarding the
effect of transport facilitators such as PIN is that they add a linear
term to the model. This turns out to be consistent with the
chemiosmotic theory, as we show in
Appendix~\ref{sec:chemiosmotic-active-transport}, so we incorporate
this idea into our Helmholtz Model of \cite{plos-paper1} as follows:
\begin{equation}
  \label{eq:dcdt-polartransport}
  \frac{\del c(i)}{\del t} = \underbrace{\sum_{j\in Nbr(i)} D
    (c(j)-c(i))}_\mathrm{Diffusion~Transport} 
  +\underbrace{\sum_{j\in Nbr} p^{in}_{j} c(j) - \sum_{j\in Nbr}
    p^{out}_j c(i)}_\mathrm{Polar~Transport} 
  + \frac{K}{S(i)} -
  \alpha c(i) ~.
\end{equation}

\noindent
Here $p^{in}_j$ is the (active) polar transport coefficient for auxin
being `pushed' into the current cell $i$ from a neighboring cell $j$,
while $p^{out}_j$ is the polar transport coefficient for auxin being
`exported' from the current cell toward the neighbor $j$. Since these
coefficients represent the asymmetrically located putative carrier
PIN1, they also represent a measure of the work that the carrier
performs. In particular, if PIN1 works at saturation, then the $p$'s
are proportional to PIN1 concentration at an interface. Thus, the
biological appearance of PIN1 translates into the appearance of
non-zero $p$'s in this model. But how can we predict when that
happens? The Constant Gradient Hypothesis offers an answer,
provided that we can solve a non-linear transport problem.

\subsection{Formulation of Non-linear Transport Problem}



Suppose the cell sizes in the domain of interest (e.g., a young leaf)
are known and that we also know which cells express PIN1. We are given
the interfaces where the carrier is present and the direction in which
it facilitates auxin transport, but we do not know how much protein is
present there.\footnote{Note that our previous model
  \cite{plos-paper1} can make this sort of prediction. Also, the
  distribution of PIN1 can be obtained from experiments such as those
  in \cite{berleth:bipolarcell}.} PIN1 may work either in the
direction of diffusive flow or against it: these are two distinct
modes of operation. We shall see how to assign an operation mode to
each interface in the following section. Here we discuss a
mathematical tool that allows us to compute the auxin concentration at
equilibrium, $c$, assuming the modes of operation are known
everywhere. Finally, this $c$ is used in conjunction with
\refeq{eq:dcdt-polartransport} to estimate the amount of PIN1 present
at each interface.

In the first mode of operation PIN1 maintains $\Delta c < \tau_1$
because it works with diffusion. We define the following non-linear
transport function:
$$
\phi^{nl}(\Delta c) = \left\{ 
\begin{array}{lr}
D& \Delta c < \tau_1 \\
D_{fast} & \Delta c \geq \tau_1 
\end{array}
\right.~.
$$

\noindent
where $D_{fast} \gg D$. Note that the usual transport function due to
Fickian diffusion is $\phi^{Fick} (\Delta c) = D \Delta c$. The only
difference is that the diffusion coefficient increases non-linearly
when the difference in auxin concentration through the interface
exceeds the threshold $\tau_1$. The dynamics of cell $i$ become:
$$
\frac{ d c}{ d t} (i) = \sum_{j\in Nbr(i)} \phi(c(j)-c(i)) +
\frac{K}{S(i)} - \alpha c(i)
$$

\noindent
where $Nbr(i)$ denotes the neighboring cells of cell $i$, and the flux
$\phi$ now depends on whether the interface contains PIN1 helping
diffusion or not, i.e. it is either $\phi^{nl}$ or $\phi^{Fick}$.  The
effect of this formulation is that an interface with transport
function (describing the flux of auxin) $\phi^{nl}$ will have a
$\Delta c < \tau_1$ or $\Delta c \approx \tau_1$ at equilibrium for a
sufficiently large $D_{fast}$. This is equivalent to what the Constant
Gradient Hypothesis postulates for PIN1 working with
diffusion.\footnote{Note, however, that the non-linear dynamics do not
  say anything about how the system behaves outside of
  equilibrium. This is simply meant as tool to compute the
  equilibrium.}

When the carrier works against the concentration gradient, however,
the hypothesis implies that the $\abs{\Delta c}$ is maintained below
but near $\tau_2$. For the purposes of the present paper, this
behavior can be achieved by a constant flux. For example, if PIN1 is
present at the interface between cells $i$ and $j$, facilitates transport
of auxin from $i$ to $j$, and $c(i) < c(j)$ (i.e. PIN1 works against
diffusion), then the flux due to PIN1 is a constant $p_{uphill}$. In
symbols,
$$
\begin{array}{rcl}
\frac{ d c}{ d t} (i) &=& \paren{ \sum_{k\in Nbr(i)} \phi(c(k)-c(i)) +
\frac{K}{S(i)} - \alpha c(i) } - p_{uphill} \\
\frac{ d c}{ d t} (j) &=& \paren{ \sum_{k\in Nbr(j)} \phi(c(k)-c(j)) +
\frac{K}{S(j)} - \alpha c(j) } + p_{uphill} \\
\end{array}
$$

So, we only assume values for uphill polar transport and we can solve
the non-linear system to obtain the distribution of auxin
concentration at equilibrium (details in
\refsec{sec:solve-non-lindyn}).  Using this distribution we can obtain
the polar transport contribution in all cells (details in
\refsec{sec:solving-for-pin}) and obtain all the parameters of the
polar transport dynamics at equilibrium
(\refeq{eq:dcdt-polartransport} with $\frac{d c}{d t}=0$). In
particular, we obtain the values of the polar transport coefficients
$p^{in}$ and $p^{out}$ which are related to the density of PIN1
at the respective membranes.

The above analysis, together with our Schema 1 fleshed
out as the appearance of PIN1, allows us to make much more detailed
predictions. These predictions follow in \refsec{sec:prediction-veins}
and \refsec{sec:embryo}.

\subsection{Solving the Non-linear Dynamics Problem}
\label{sec:solve-non-lindyn}

Now we show how to find the equilibrium concentration of auxin
according to the non-linear transport formulation. We need to find
$c(i)$ for all cells $i$ that satisfies:
\begin{equation}
  \label{eq:non-lin}
  \frac{ d c}{ d t} (i) = \sum_{j\in Nbr(i)} \phi(c(j), c(i)) +
  \rho(i) - \alpha c(i) = 0
\end{equation}

\noindent
where $Nbr(i)$ denotes the neighboring cells of cell $i$; $\phi$ is
the flux of auxin from cell $j$ into cell $i$;
$\rho(i)=\frac{K}{S(i)}$ is the production function; and $\alpha$ is
the destruction constant. There are three types of flux functions of
interest: diffusion according to Fick's Law $\phi^{Fick}$; non-linear
diffusion $\phi^{nl}$; and polar transport $\phi^{polar}(c(j),
c(i))=\pm p_{uphill}$ where the sign is positive if PIN1 facilitates
transport uphill from cell $j$ into cell $i$ and negative in the
opposite direction. All three share properties that guarantee the
dynamical system to have a unique solution.

\begin{prop}
  Let $\phi(c(i),c(j))$ have the following properties:
  \begin{enumerate}
    \item Antisymmetry: $\phi(x,y) = - \phi(y,x)$

    \item Monotonicity: $x-y \leq v-w \implies
      \phi(x,y) \leq \phi(v,w)$

    \item $\phi_x = \frac{\partial \phi}{\partial x}\geq 0$.
  \end{enumerate}

  Then, there is a unique $c$ that satisfies \refeq{eq:non-lin} which
  can be found using a Newton method.
\end{prop}

\begin{proof}

Let $c$ be the distribution of concentration and define the dynamics
as:
$$ 
(c_t)(i) = \sum_{j\in Nbr(i)} \phi(c(j), c(i)) +\rho(i) - \alpha
c(i) ~.
$$

Now the Jacobian with respect to $c(i)$ is a matrix 
$$
J = \left[ 
  \frac{\partial (c_t)(i)}{\partial c(j)}
\right]_{ij}
$$ 

\noindent
with off-diagonal entries, $i\neq j$, $J_{ij}=\frac{\partial
  \phi(c(j),c(i))}{\partial c(j)} = \phi_x(c(j),c(i))$. The diagonal
entries are
$$ 
J_{ii}=\frac{\partial (c_t)(i)}{\partial c(i)} = \sum_{j\in
  Nbr(i)} \frac{\partial \phi(c(j), c(i))}{\partial c(i)} - \alpha~=
\sum_{j\in Nbr(i)} \phi_y(c(j), c(i)) - \alpha
$$

\noindent
where $\phi_y$ denotes the partial derivative with respect to the
second argument.

Now, using the antisymmetry property, differentiating with respect to
$x$ on both sides, we see that $\phi_x = -\phi_y$. Therefore, the
diagonal entries of the Jacobian become
$$ 
J_{ii} = \sum_{j\in Nbr(i)} -\phi_x(c(j), c(i)) - \alpha~.
$$

\noindent
Thus, the third property implies that the diagonal entries are
strictly negative and that they dominate the off-diagonal entries
(because $\sum_{j\neq i} J_{ij} = \sum_{j\in Nbr(i)} \phi_x(c(j),
c(i))$). As a result, the Jacobian $J$ is always invertible and
Newton's method applies. In particular, this means that $J (c_t) = 0
\iff (c_t) = 0$ so such a $c$ exists because the $c>0$ and $\sum_i c
\leq \sum_i \rho/\alpha$ under these dynamics.

It is also unique. Let the transport operator $\Phi$ be defined as 
$$ 
(\Phi c)(i) = \sum_{j\in Nbr(i)} \phi(c(j), c(i))
$$

\noindent
and suppose that both $x$ and $y$ are solutions. Thus, 
$$
\Phi x - \alpha x = -\rho = \Phi y - \alpha y
\implies \Phi x - \Phi y = \alpha (x - y)
$$

\noindent
So the inner product 
\begin{equation}
  \label{eq:inner-neg}
  (\Phi x - \Phi y, x-y) = \alpha (x-y,x-y)\geq
  0~.
\end{equation}

\noindent
On the other hand
\begin{equation}
  \label{eq:innerprod1}
  (\Phi x - \Phi y, x-y) = (\Phi x, x) - (\Phi x, y) - (\Phi y, x) +
  (\Phi y, y)
\end{equation}

\noindent
where 
$$
 (\Phi x, x) = \sum_i x_i (\Phi x)(i) = \sum_i x_i \paren{\sum_{j\in
  Nbr(i)}\phi(x_j, x_i) }~.
$$
\noindent
Each pair of neighbors appears exactly twice in the flux function:
once for each direction. Thus, the term $x_i \phi(x_j, x_i)$
corresponds to the term $x_j \phi(x_i, x_j)$. Therefore, using the
first property we can pair the two terms as $(x_i-x_j) \phi(x_j,
x_i)$. The expression for the $(\Phi x,x)$ then becomes
$$
(\Phi x, x) = \sum_{j\neq i~Nbrs} (x_i-x_j) \phi(x_j,
x_i)
$$
\noindent
where the sum is over all pairs of neighbors $i$ and $j$. Using the
same argument, we obtain expressions for the other inner products and
rewrite \refeq{eq:innerprod1} as
$$
\begin{array}{rl}
  (\Phi x - \Phi y, x-y) \\
=& \sum (x_i-x_j) \phi(x_j, x_i)
  - (y_i-y_j) \phi(x_j, x_i) - (x_i-x_j) \phi(y_j, y_i) + (y_i-y_j)
  \phi(y_j, y_i) \\
 =& \sum -(x_i-x_j) \phi(x_i, x_j)
  + (y_i-y_j) \phi(x_i, x_j) + (x_i-x_j) \phi(y_i, y_j) - (y_i-y_j)
  \phi(y_i, y_j)
\end{array}
$$

\noindent
Let $A=x_i-x_j$, $\phi(A)=\phi(x_i,x_j)$, $B=y_i- y_j$, and $\phi(B)=\phi(y_i,
y_j)$. So we can rewrite the terms in the summation as
$$
-A \phi(A) + B \phi(A) + A \phi(B) - B\phi(B) = -(A-B)(\phi(A)-\phi(B))~.
$$

\noindent
The second property of $\phi$ guarantees that $(A-B)$ and
$\phi(A)-\phi(B)$ have the same sign,
i.e. $-(A-B)(\phi(A)-\phi(B))\leq 0$. Therefore, $(\Phi x - \Phi y,
x-y) \leq 0$. Thus, recalling \refeq{eq:inner-neg}, we see that $(\Phi
x - \Phi y, x-y) = 0$, which either means that $x=y$ and we are done,
or that $\Phi x = \Phi y$. In the latter case, we have
$$
\Phi x -\alpha x= \Phi y - \alpha y \implies 
-\alpha x=  - \alpha y \implies x=y
$$

\noindent
so the solution is unique.
\end{proof}

\subsection{Computing the Polar Transport Contribution}
\label{sec:solving-for-pin}

Now suppose that the distribution of auxin concentration at
equilibrium, $c$, is known---e.g., computed as in
\refsec{sec:solve-non-lindyn}---and that the polar transport dynamics
are given by
\begin{equation}
  \label{eq:polar-dyn} 
  \frac{ d c}{ d t} (i) = \sum_{j\in Nbr(i)} \paren{D(c(j)-c(i)) -
    \pi_{i\to j}+\pi_{j\to i} } + \frac{K}{S(i)} - \alpha c(i) = 0
\end{equation}

\noindent
where $\rho(i)=K/S(i)$ and $\alpha$ are also known. Our task here is
to compute the effective polar transport contribution: the $\pi_{i\to
  j}$. The unsaturated polar transport and the effective polar transport
are related
\begin{equation}
  \label{eq:TPT}
  \pi_{i\to j} = p_{i\to j} c(i)
\end{equation}

\noindent
so we can rewrite \refeq{eq:polar-dyn} in matrix form:
\begin{equation}
  \label{eq:polar-dyn-matrix} 
  \mathcal{D} \v{c} + \mathcal{P}\v{c} + \rho - \alpha \mathcal{I} \v{c}=
0
\end{equation}

\noindent
where $\mathcal{D}$ is the diffusion matrix as described in
\cite{plos-paper1}; $\mathcal{P}$ is the polar transport matrix;
$\rho$ is the production vector; and $\mathcal{I}$ is the identity
matrix. The polar transport matrix is everywhere zero except for
interfaces $i|j$ with polar transport. Thus, if the direction of
transport is $i\to j$, then $\mathcal{P}_{ij}=p_{i\to j}$ and
$\mathcal{P}_{ii}=-p_{i\to j}+Other$ (where $Other$ refers to other
polar interfaces that cell $i$ may have). This contributes $-p_{i\to
  j} c(i)$ to the dynamics of cell $i$ and $p_{i\to j} c(i)$ to the
dynamics of cell $j$ as required. The unknowns are the $p_{i\to j}$,
i.e. the entries of $\mathcal{P}$. Let $\v{p}$ be the vector of those
unknowns and write
\begin{equation}
  \label{eq:syseq-pin}
  (\mathcal{P}^*) \v{p} = \mathcal{P}\v{c}~.
\end{equation}

\noindent
\refeq{eq:polar-dyn-matrix} implies that 
\begin{equation}
\label{eq:Pceq}
\mathcal{P}\v{c} =
(-\mathcal{D} + \alpha \mathcal{I}) \v{c} - \rho,
\end{equation}
which contains no unknowns and can be computed. The matrix
$\mathcal{P}^*$ is the dual of $\mathcal{P}$ and can be written as
follows. The $i^{th}$ coordinate of the vector $\mathcal{P}\v{c}$
consists of the sum of the contribution of all outgoing polar
transport activity---cell $i$ exporting to neighbors---and all
incoming activity---neighbors exporting toward cell $i$.
\begin{equation}
  \label{eq:P*init}
        [(\mathcal{P}^*) \v{p} ]_i  = \sum_{j\in out}-p_{i\to
          j} c(i) + \sum_{j\in in} p_{j\to i} c(j) = [\mathcal{P}\v{c}]_i
\end{equation}

\noindent
so combining with \refeq{eq:TPT} we obtain an expression $\pi_{i\to j}$
\begin{equation}
  \label{eq:P*}
        [\Pi \vec{\pi} ]_i = \sum_{j\in out}-\pi_{i\to j} +
        \sum_{j\in in} \pi_{j\to i} = [\mathcal{P}\v{c}]_i
\end{equation}

\noindent
where $\vec{\pi}$ is the vector of effective polar transport
variables, and $\Pi$ is the $n\times m$ incidence matrix for the polar
transport graph where $n$ is the number of cells and $m$ is the number
of edges. Thus, letting $k$ denote the edge $e_k = (i\to j)$, we have
$$
\Pi_{lk} = \left\{ 
\begin{array}{rl}
+1, & \mathrm{if}~ l = j \\
-1, & \mathrm{if}~ l = i \\
0, & \mathrm{otherwise} .
\end{array}
\right.
$$ The polar transport graph is a subgraph of the cell graph and
consists only of those cells which are incident to an edge that
corresponds to an unknown $\pi_{i\to j}$---there are $n'\leq n$ such
cells. In particular, only $n'$ out of the $n$ rows of $\Pi$ have
non-zero entries.

\begin{prop}
We can compute $\vec{\pi}$---and therefore $\v{p}$---if and only if the
polar transport graph is an undirected forest, i.e. $m = n'-c$ ($c$ is
the number of connected components), and $\v{c}$ was obtained from a
transport rule $\phi$ that satisfies $\phi(i\to j) = -\phi(j\to i)$
for each edge $(i,j)$.
\end{prop}

\begin{proof}
  First, we note that the rank of $\Pi$ is $n'-c$ by Proposition 4.3
  in \cite[p.~24]{briggs93}. Thus, the non-zero portion of $\Pi$ is
  square and invertible if and only if $m = n'-c$. Suppose that
  $\vec{\pi}$ was computed using this square matrix. The variables
  satisfy all $n'$ equations, not only the $n'-c$. To see why, note
  that $\pi_{i\to j}$ appears in exactly two equations but with
  opposite signs. Also, diffusive transport works in the same way, as
  did the original transport rule. Thus, both sides of \refeq{eq:Pceq}
  satisfy the property and $\vec{\pi}$ is a solution to the whole
  system \refeq{eq:P*}.
\end{proof}

The exact nature of transport facilitation that PIN1
provides is unknown, but the effective polar transport provides a
measure of the effect. Hence, assuming a model of auxin transport by
PIN1, we can compute how much protein is needed to achieve the effect
through a relation analogous to \refeq{eq:TPT}. In the simplest model
the density of PIN1 is only proportional to $\pi$.

\section{Predictions of Vein Patterns in Leaves}
\label{sec:prediction-veins}

\begin{figure}  
  \begin{center}
    \begin{tabular}{cc}
      \includegraphics{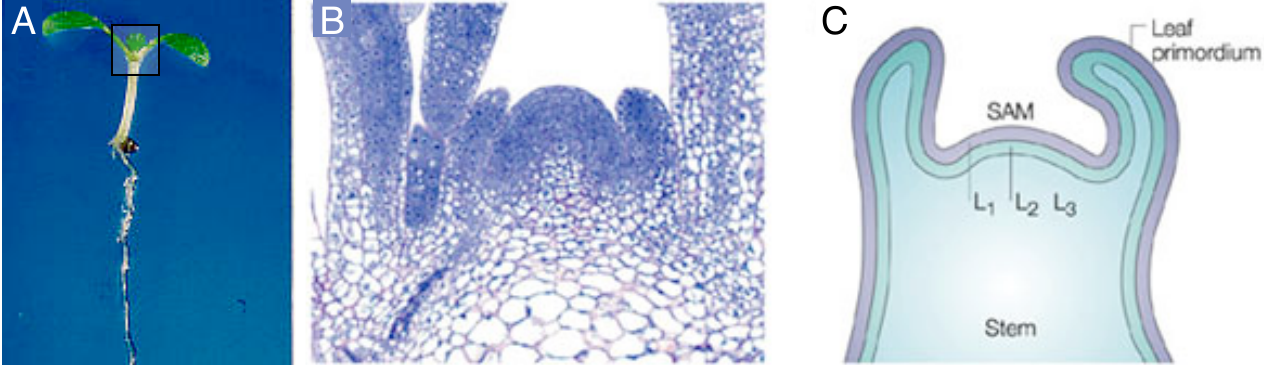}
    \end{tabular}
  \end{center}
  \caption[Structure of shoot apex and location of leaf
    primordium.]{\label{fig:pin-athalianasam-1} Structure of shoot
    apex and location of leaf primordium. \FP{A} {\em Arabidopsis
      thaliana} seedling (from \cite{olivier-etal-2006}). \FP{B} A
    slice through the shoot apex of {\em A. thaliana}
    \cite[Fig.~5]{lucas-and-lee-2004} found under highlighted area in
    \CP{A}. \FP{C} Schematic of slice in \CP{B}. SAM: shoot apical
    meristem, responsible for growth along the main axis. }
\end{figure}

\begin{figure}
  
  \begin{center}
    \begin{tabular}{cc}
      \includegraphics{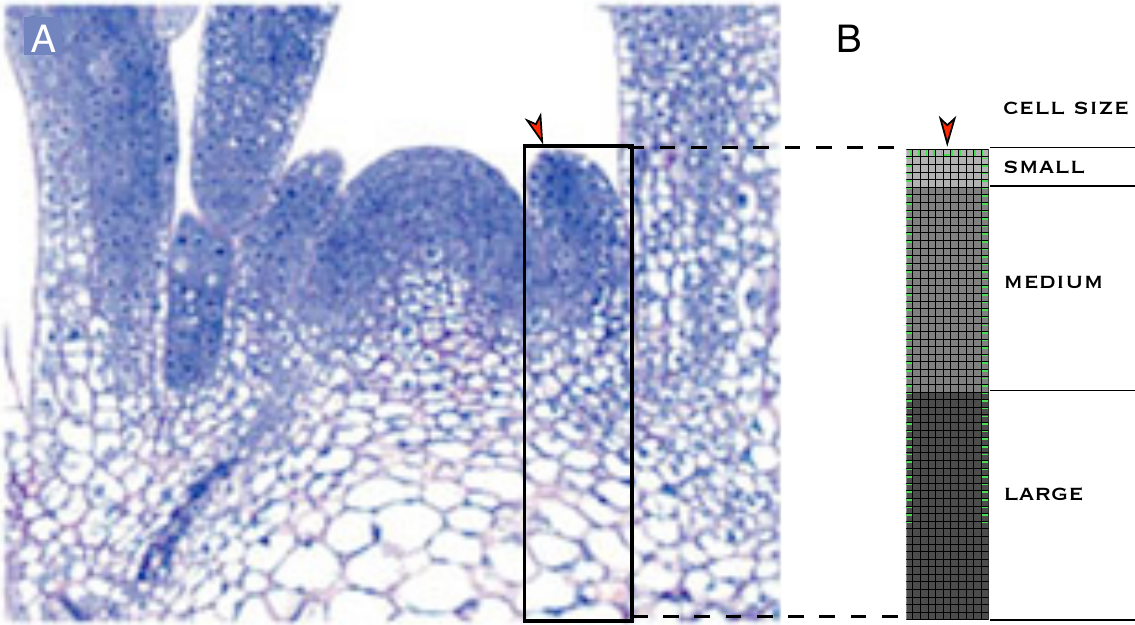}
    \end{tabular}
  \end{center}
  \caption[Setup for simulations.]{\label{fig:pin-veinsim-setup-1}
    Setup for simulations. \FP{A} Shoot apex of {\em A. thaliana} as
    in \reffig[C]{fig:pin-athalianasam-1}. \FP{B} Simulation domain
    corresponding to region of real organ. Red arrowhead points at tip
    of leaf primordium. Cell sizes increase moving away from the tip,
    compare with \CP{A} where cells are largest in the stem.}
\end{figure}

We shall now show that we can not only predict where and when new
c-vascular strands form, but also the relative strength of polar
transport that implements the required improvement in auxin
transport. The the best of our knowledge, PIN1 is a likely and,
possibly, paradigmatic instance of molecular polar transport and a
precursor to the venation pattern. However, the pattern of PIN1
expression, referred to as PED (or PIN1 expression domain), does not
correspond exactly to the vein pattern. Recent experimental studies by
Scarpella~\etal~\cite{berleth:bipolarcell} demonstrate that only a
subpart of the PED becomes vein pattern. Surprisingly, some parts of
the PED disappear. Here, we shall show that our theory is consistent
with these observations by showing how each of the major empirically
defined stages can be explained within our framework. We shall model
the behavior of cells within a developing leaf---the so-called leaf
primordium---and the adjoining cells of the shoot apex (see
\reffig{fig:pin-athalianasam-1} and
\reffig{fig:pin-veinsim-setup-1}). We begin with a cartoon domain of
cells, to show qualitative properties, and then proceed to compare
these results to experimental data.

\subsection{Emergence of leaf primordium and formation of midvein}
\label{sec:midvein-emergence}

\begin{figure}
  \begin{center}
    \begin{tabular}{cc}
      \includegraphics{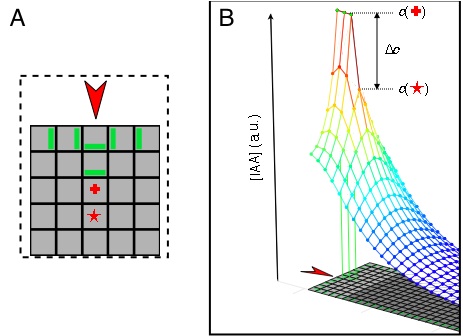}
    \end{tabular}
  \end{center}
  \caption[Predictions of auxin concentration during initial stages of
    primordium formation.]{\label{fig:pin-veinsim-setup-2} Predictions
    of auxin concentration during initial stages of primordium
    formation. \FP{A} Magnification of the tip of our simulation
    domain. Carrier expression localizes asymmetrically and, along the
    marginal cells, points toward a convergence point---cell with red
    arrowhead. \FP{B} The predicted distribution of auxin. Note that
    the largest $\Delta c$ is between the cell marked with a red cross
    and its neighbor marked with a red star. This is the most likely
    place for new PIN1. As \CP{A} shows, these cells are at the tip of
    the inward forming midvein PIN1 expression domain.  Exaggerated
    $c$ for display purposes.}
\end{figure}

Our model predicts the emergence of a midvein as a result of leaf
growth. The formation of a new leaf primordium on the shoot apical
meristem (SAM, see \reffig{fig:pin-athalianasam-1}) begins by the
reorientation of polarity in some cells and the formation of
``convergence points''---red arrowheads in
\reffig{fig:pin-veinsim-setup-2}. This, as our model predicts, results
in the creation of a locally extreme difference in auxin
concentration. \reffig[B]{fig:pin-veinsim-setup-2} illustrates that
$\Delta c$ is largest near the convergence point between the cell
marked with a cross and its neighbor marked with a red star. The large
$\Delta c$ promotes PIN1 creation and, eventually, the interface on
the side of the cell with a cross is endowed with the
carrier. Following the extension of the middle PED, the extreme
$\Delta c$ moves to the next interface of the cell with the star and
the procedure repeats. But each time the PED is extended in this
fashion, the extreme $\Delta c$ decreases until it is too small to
create any new PIN1 (\refsec{sec:pin-polarity}).

\begin{figure}
  \begin{center}
    \begin{tabular}{cc}
      \includegraphics{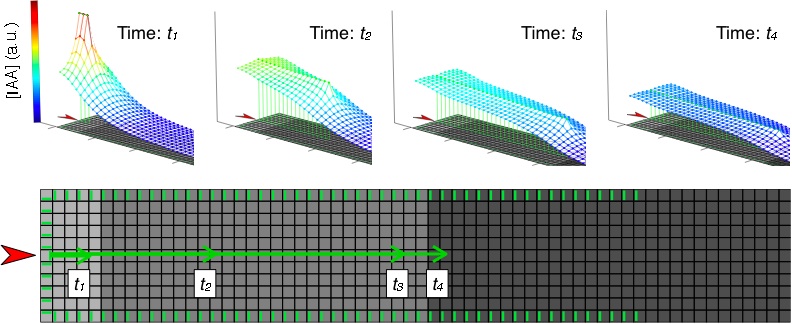}
    \end{tabular}
  \end{center}
  \caption[Prediction of midvein from MPED using
    \refalg{alg:midvein-from-mped}.]{\label{fig:pin-veins-midvein-1}
    Prediction of midvein from MPED using
    \refalg{alg:midvein-from-mped}. The domain and the marginal PIN1
    expression, green bars, are shown. Also shown are four snapshots
    during the execution of \refalg{alg:midvein-from-mped} labelled
    $t_1,t_2,t_3,$ or $t_4$. The new PED forms where the midvein of
    the new leaf should be, green arrows. Note that this pattern is
    not prescribed, but the result of implementing our Schema 1 of
    \cite{pdswz:paper} as \refalg{alg:midvein-from-mped}. }
\end{figure}

To illustrate, consider \reffig{fig:pin-veins-midvein-1}. We are given
the marginal PIN1 expression and follow the predictions of our model
through time as they give rise to the midvein. At time $t_1$ we show
the configuration from \reffig{fig:pin-veinsim-setup-2}. Then new PIN1
appears and creates a large $\Delta c$ at the tip of the PED. Observe
that this $\Delta c$ is smaller at time $t_2$ than at time $t_1$. It
is even smaller at time $t_3$ and, at time $t_4$, it becomes too small
to create new PIN1. So the midvein PED extends a finite distance from
the primordium tip toward the stem below.

\begin{figure}
  \begin{center}
    \begin{tabular}{cc}
      \includegraphics{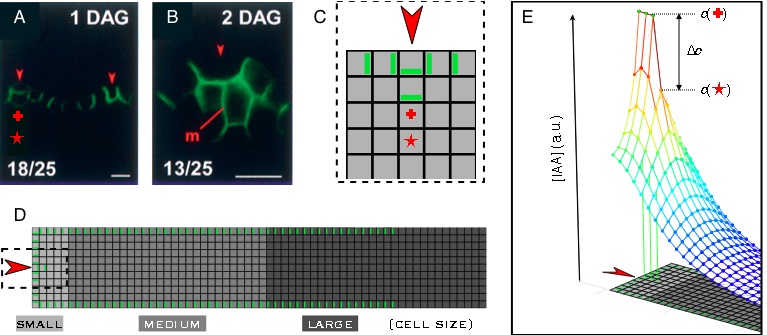}
    \end{tabular}
  \end{center}
  \caption[Predictions of auxin concentration during initial stages of
    primordium formation.]{\label{fig:pin-veinsim-comparison-1}
    Predictions of auxin concentration during initial stages of
    primordium formation. \FP{A,B} Initial stages in the formation of leaf
    primordia (from \cite{berleth:bipolarcell}). Sub-cellular
    localization of PIN1 directed toward so-called ``convergence
    points'': cells shown with red arrowheads. \FP{C} Model of this
    behavior. PIN1, depicted in green, organize in marginal PEDs
    (MPED) facilitating auxin flow toward the tip of the primordium
    (the convergence point) from all sides. The hormone is evacuated
    by the inwardly directed PIN1 expression that will eventually
    become part of the midvein in the new leaf. \FP{D} Entire
    simulation domain. Selected region enlarged in \CP{C}. \FP{E}
    Distribution of auxin concentration at equilibrium computed as
    described in \refsec{sec:non-linear-model}. The midvein PED works
    with diffusion---mode 1---and the MPEDs work against
    diffusion---mode 2. The largest $\Delta c$ is at the end of the
    midvein PED between the cell marked with a cross and the neighbor
    marked with a red star. This is the most likely place for new
    PIN1. Exaggerated $c$ for display purposes.}

\end{figure}

\begin{figure}
  \begin{center}
    \begin{tabular}{cc}
      \includegraphics{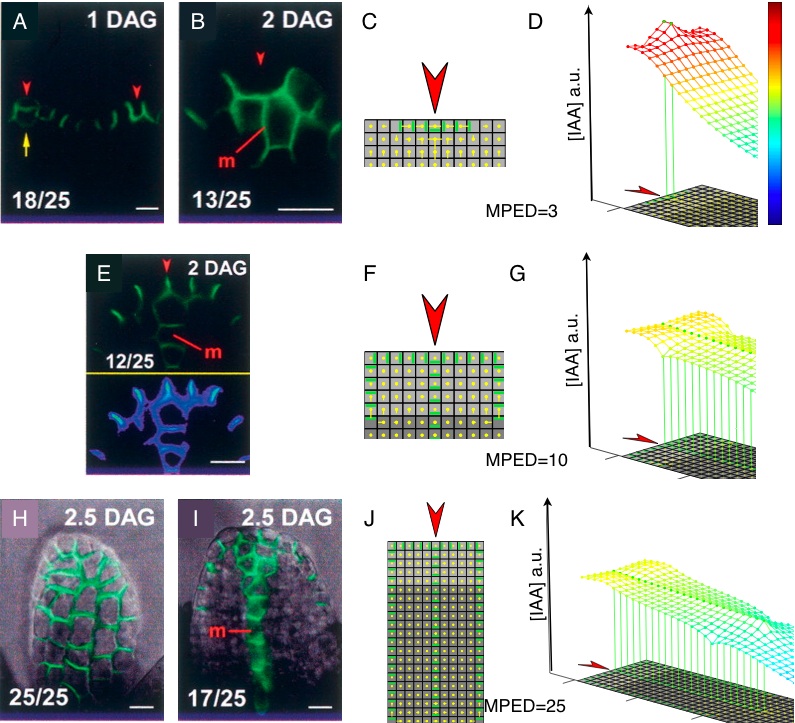}
    \end{tabular}
  \end{center}
  \caption[Growth of leaf primordium induces growth of
    midvein.]{\label{fig:pin-veins-midvein-2} Growth of leaf
    primordium induces growth of midvein. Note that as the midvein
    grows, it connects the tip of the leaf---where cells are small so
    $c$ is high---to cells in the stem---which are large so $c$ is
    low. Thus, the concentration at the tip of the leaf primordium
    decreases. In effect, the midvein acts as a straight bar being
    pushed up at the tip but pulled down near the stem. This is due to
    PIN1 following Constant Gradient Hypothesis. Three
    primordium sizes are shown in increasing order: \CP{A--D},
    \CP{E--G}, and \CP{H--K}. \FP{A,B,E,H,I} from
    \cite{berleth:bipolarcell}. }
\end{figure}

\begin{figure}
  \begin{center}
    \begin{tabular}{cc}
      \includegraphics{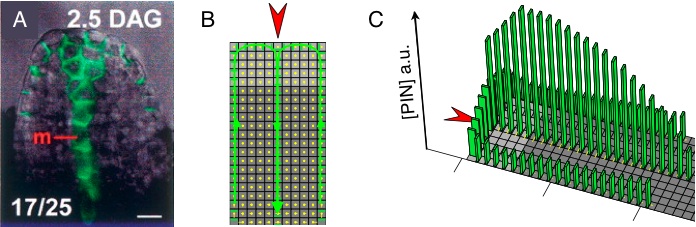}
    \end{tabular}
  \end{center}
  \caption[Prediction of PIN1
    strength.]{\label{fig:pin-veins-midvein-pinstrength} Prediction of
    PIN1 strength computed as described in
    \refsec{sec:non-linear-model}. \FP{A} Sample leaf from
    \cite{berleth:bipolarcell}. PIN1 expression in green, stronger
    just under tip. \FP{B} PIN1 polarity shown as green bars. Flow due
    to polar transport shown as green arrows. \FP{C} Prediction of
    PIN1 concentration. Height of green bars denotes
    concentration. The expression is low along the margin and high in
    the midvein for both the real leaf, in \CP{A}, and predictions, in
    \CP{C}. }
\end{figure}

In fact, these predictions are consistent with experimental
observations. Our initial setup mimics the biology,
\reffig{fig:pin-veinsim-comparison-1}, and our predictions for an
extending midvein during leaf growth coincide with the empirical
patterns \reffig{fig:pin-veins-midvein-2}. But our predictions go even
further. We compute the relative demand for polar transport along the
whole domain under the Constant Gradient
Hypothesis. Thus, if the concentration of PIN1 is
proportional to this demand we predict the distribution of the carrier
throughout the leaf. \reffig{fig:pin-veins-midvein-pinstrength}
provides one example. Note that more auxin carrier is needed in the
midvein than in the marginal PEDs, which is consistent with PIN1
expression in a real leaf.

\subsection{Convergence points on leaf margin}

The pattern of PIN1 expression shown in
\reffig{fig:pin-veins-midvein-pinstrength} is maintained as the young
leaf grows. In general, when a cell expressing PIN1 divides the
daughter cells inherit the polarity of the mother cell. This rule is
broken only occasionally, and when that happens a so-called
convergence point results. Our model can explain this phenomenon in
the following fashion.

\begin{figure}
  \begin{center}
    \begin{tabular}{c}
      \includegraphics{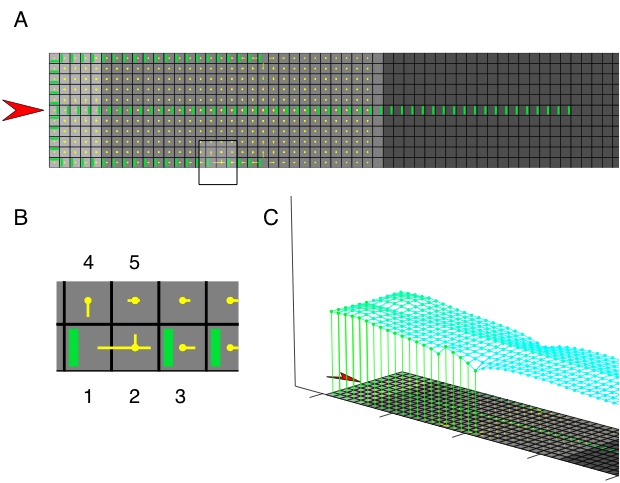}
    \end{tabular}
  \end{center}
  \caption[Auxin distribution following a cell division on the leaf
    margin.]{\label{fig:celldiv-margin} Auxin distribution following a
    cell division on the leaf margin. \FP{A} The domain. Rectangular
    area blown up in \CP{B}. \FP{B} $\Delta c$ shown as a yellow
    segment starting from the cell with the higher
    concentration. E.g., the largest $\Delta c=c(2)-c(1)$. Since PIN1
    is created due to large differences in concentration and in the
    direction of that concentration, the new PIN1 accumulates inside
    cell 2 toward cell 1. Polarity is maintained consistent with
    neighboring cells. \FP{C} Auxin concentration shown as a height
    map over the domain.}
\end{figure}

First, examine what happens when a cell on the margin
divides. \reffig{fig:celldiv-margin} illustrates the distribution of
auxin concentration after the new cell membrane (or wall) has formed
sufficiently to present a barrier and the concentrations do not
change. Notice, in \reffig[B]{fig:celldiv-margin}, that the largest
difference in concentration is between the two daughter cells and that
it suggests that new PIN1 should appear so that the polarity is
consistent with neighboring cells. This behavior is generic ---
\reffig[C]{fig:margin-flip} demonstrates it for all cells along the
margin but it also holds in the midvein.

\begin{figure}
  \begin{center}
    \begin{tabular}{c}
      \includegraphics{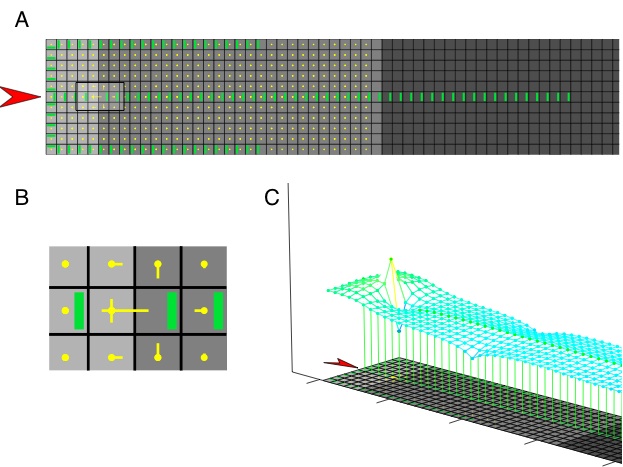}
    \end{tabular}
  \end{center}
   \caption[Auxin distribution following a cell division at the
     midvein PED.]{\label{fig:celldiv-midvein} Auxin distribution
     following a cell division at the midvein PED. \FP{A} The
     domain. Rectangular area blown up in \CP{B}. \FP{B,C} Largest
     $\Delta c$ through new interface as in
     \reffig{fig:celldiv-margin}.}
\end{figure}

The situation in the midvein is somewhat different because there are
more neighbors and the strength of PIN1 is largest in the
midvein. Still, \reffig{fig:celldiv-midvein} shows that PIN1 polarity
can be maintained in the midvein for much the same reasons as
before. However, the ensuing difference in concentration through the
newly formed interface is significantly larger for a dividing midvein
cell than for a dividing margin cell, \reffig[D]{fig:margin-flip}
demonstrates this graphically. As a result, new PIN1 should appear
quicker and have an effect on the concentration faster for dividing
cells in the midvein than for dividing cells along the
margin. Therefore, any PIN1 that may start forming at interfaces other
than the new one would probably not form in sufficient quantity to
remain there after the new interface has acquired its share of
PIN1.\footnote{In other words, the fast action of PIN at new interface
  will not allow the $\Delta c$ at neighboring interfaces to exceed
  $\tau_1$ long enough to make $p > p_\tau$ at the neighboring
  interfaces. Here $\tau_1$ is such that the PIN1 dynamics can, given
  sufficient time, switch to the ``on'' position,
  i.e. $[PIN1]>p_\tau$. 
}
The cells along the leaf margin, on the other hand, cannot create
$\Delta c$ quite as large and the ensuing slower acquisition of PIN1
can result in PIN1 appearing---and persisting---at interfaces other
than the newly created one.

\begin{figure}
  \begin{center}
      \includegraphics{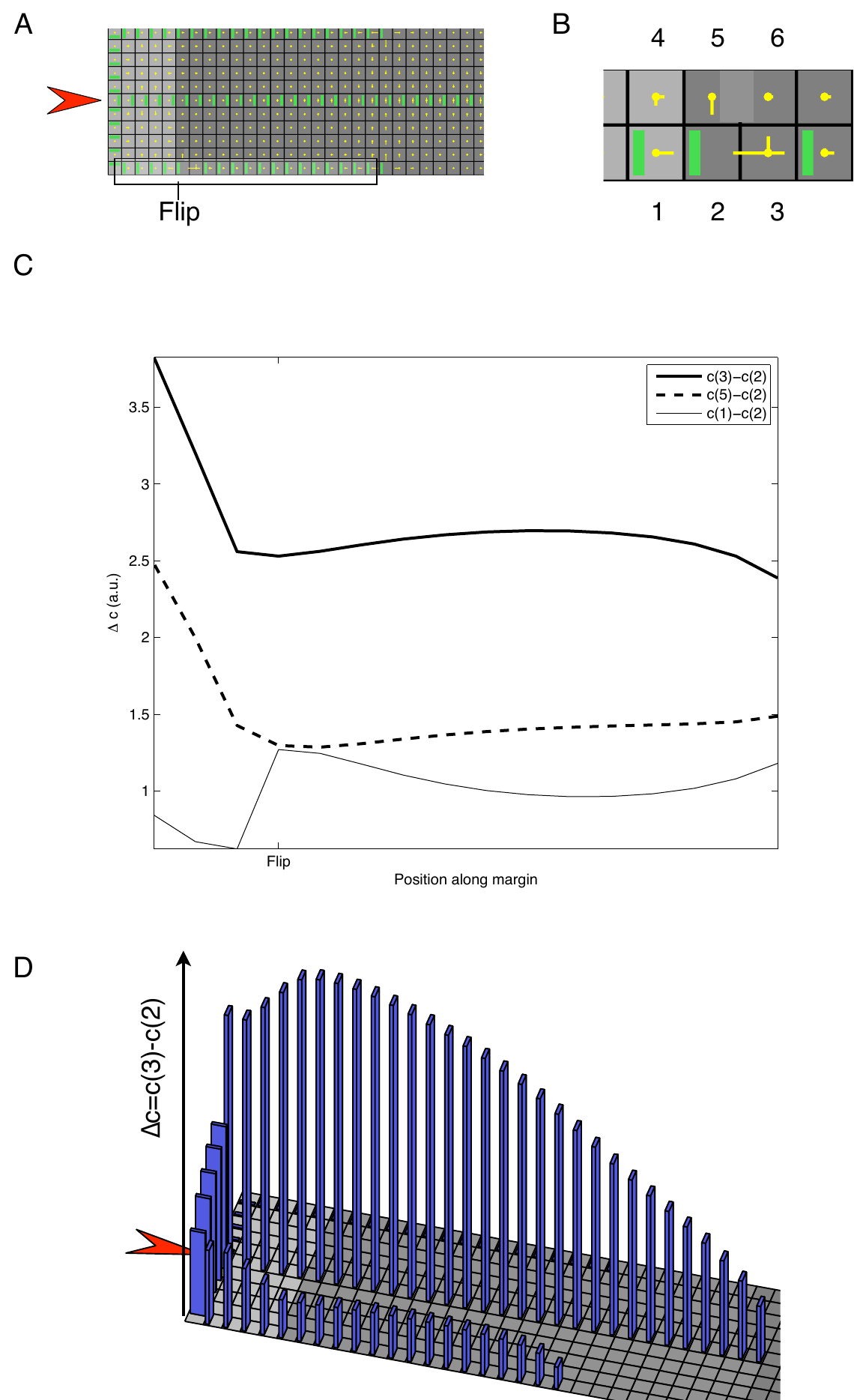}
  \end{center}
    \vspace{-1.5em}
      \caption[Comparison of cell division effects along leaf
        margin.]{\label{fig:margin-flip} Cell division effects along
        leaf margin. \FP{A} The rectangle encloses the cells used in
        \CP{C}. Example of one cell dividing while all others do
        not. \FP{B} The neighborhood of a cell that has recently
        divided. The daughter cells are labeled by the numbers 2 and
        3, and the new interface between them does not yet express
        PIN. The curves in \CP{C} are obtained by sliding this
        neighborhood along the margin (rectangle in \CP{A}): think of
        the position of cell 1 as moving along the margin, while the
        other compartments have positions relative to that of cell
        1. \FP{C} Three curves obtained from measuring $\Delta c$ as
        indicated. \FP{D} Effect of cell division along the whole
        PED. Hormone distribution computed for each location
        separately assuming that a cell has divided at that location
        only (as in \reffig{fig:celldiv-margin} or
        \reffig{fig:celldiv-midvein}). Height of bar is $\Delta
        c$ through the new interface without PIN. Aggregate
        plot.}
\end{figure}

\begin{figure}
  \begin{center}
    \begin{tabular}{c}
      \includegraphics{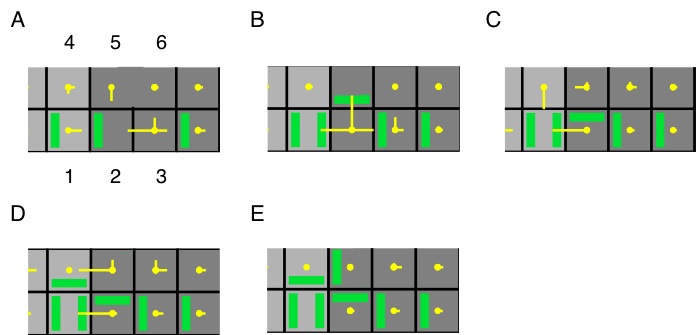}
    \end{tabular}
  \end{center}
   \caption[Creation of a convergence point at the leaf margin due to
     a cell division.]{\label{fig:bipolar-formation} Creation of a
     convergence point at the leaf margin due to a cell division. PIN1
     is created where $\Delta c$ is large in the direction of arrows;
     destroyed if against such arrows. \FP{A} Initial
     configuration. Cell divided into cells 3 and 2. Expect new
     interface to acquire PIN1 $3\to 2$. However, $\Delta c$ at $1\to
     2$ is large enough, so PIN1 will flip there. Also, $5\to 2$ will
     be created. \FP{B} The predicted configuration form \CP{A}. The
     largest $\Delta c$ is against the PIN1 acting $5\to 2$ so
     polarity will flip. \FP{C} Updated configuration from
     \CP{B}. \FP{C,D} If the supporting PIN1 can appear fast enough,
     then the $\Delta c$ through $2 \to 1$ will not eliminate the
     original flip and the configuration will be stable as in
     \CP{E}. \FP{E} Stable configuration: there are no more high
     $\Delta c$. Cell 1 is a bipolar cell because PIN expresses on
     both its left and right interfaces. Cell 2 is a convergence point
     because cells on both sides exhibit polar transport toward cell
     2.}
\end{figure}

The most interesting phenomenon of this sort is the flip of polarity
at a nearby interface, which creates a convergence point: a stable
discontinuity of PIN1 polarity along the margin.
\reffig[B]{fig:margin-flip} illustrates the differences in
concentration resulting from a cell division that created cells 2 and
3. Besides the largest $\Delta c$ through the new interface, there are
other interfaces with non-trivial $\Delta
c$. \reffig[C]{fig:margin-flip} plots three of the largest $\Delta c$
as a function of the location of the dividing cell along the
margin. Of the three, only the $\Delta c$ through interface $1\to 2$
can break the continuity of polar cells and create a ``flip''. The
most likely location for such a flip is where $\Delta c$ is largest
and it is possible for a flip to remain
there. \reffig{fig:bipolar-formation} illustrates a likely sequence of
events that results in a stable flip of polarity. In particular, this
gives rise to the so-called convergence point on the margin. The
stability of this discontinuity is due the relatively slower dynamics
of PIN1 along the margin as compared to the midvein are important
here. If a flip were to occur in the midvein, then the resulting
$\Delta c$ against the action of PIN1 will grow faster than the
neighboring interfaces can compensate and, eventually, will make PIN1
flip back. Hence, the flip will be unstable in the midvein.

\begin{figure}
  \begin{center}
    \begin{tabular}{c}
      \includegraphics{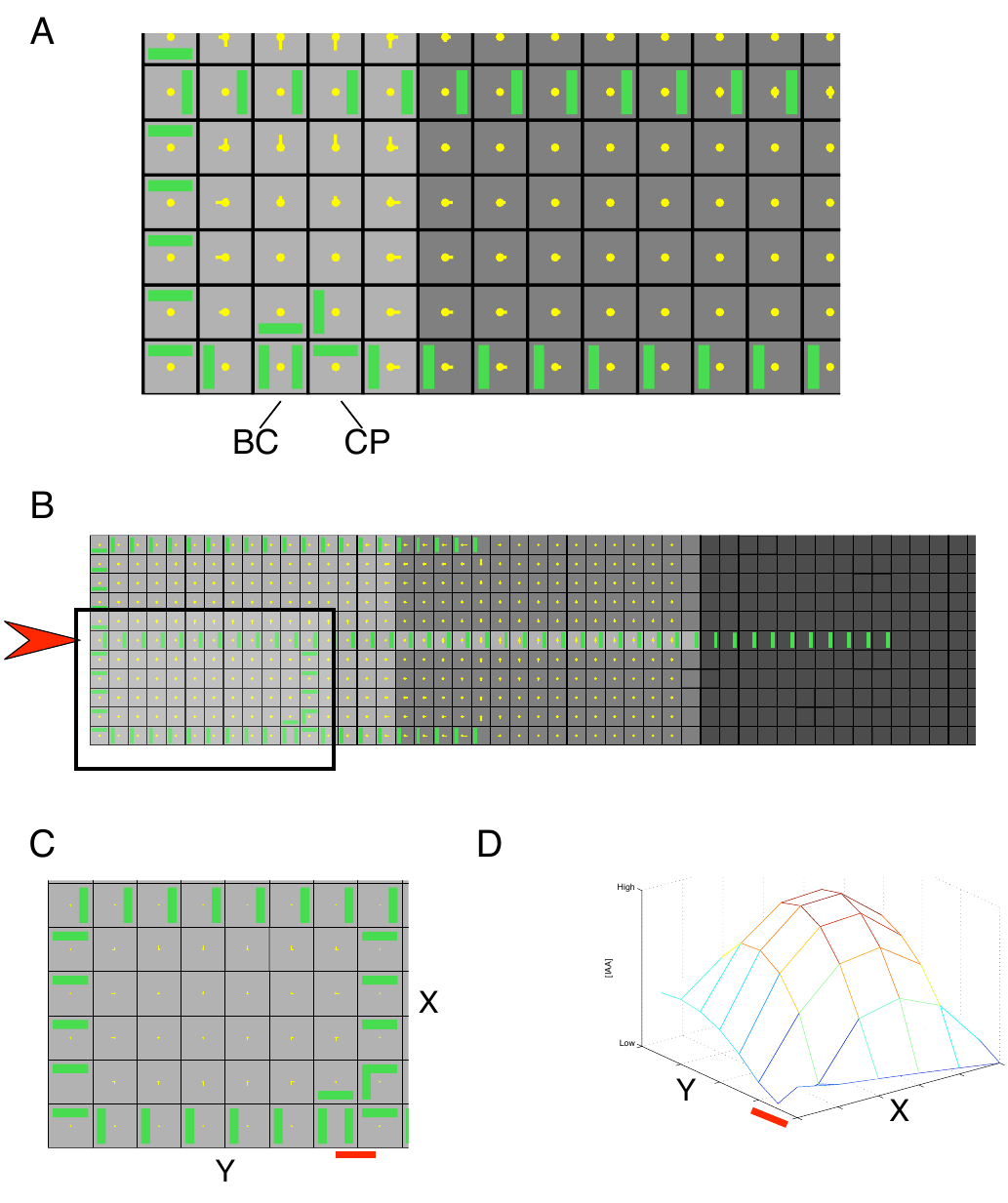}
    \end{tabular}
  \end{center}
   \caption[Emergence of strand connecting convergence point to
     midvein.]{\label{fig:margin-to-midvein} Emergence of strand
     connecting convergence point (CP) to midvein; creation of
     geometric areole. \FP{A} Distribution of $\Delta c$. Notice that
     the $\Delta c$ at the convergence point is zero so there is a
     peak of concentration there. Conversely, the largest $\Delta c$
     is at the midvein, so a new strand could emerge from the midvein
     and end at the convergence point by following the usual local
     rules. BC: bipolar cell; CP: convergence point. \FP{B} Domain
     with new strand. Rectangle encloses the geometric areole. \FP{C}
     Close up of geometric areole. Note the red bar under the
     interface between the BC and the CP. \FP{D} Concentration, $c$,
     of auxin in geometric areole. Red bar corresponds to red bar in
     \CP{C}. Note that $c(CP)>c(BC)$ so PIN1 works against diffusion
     and an auxin infusion into CP can flip the polarity of the
     interface between BC and CP.}
\end{figure}

\subsection{Formation of the first loop}

Once a convergence point has been established the pattern of PIN may
continue progressing inwardly in a manner similar to the midvein
appearance (\refsec{sec:midvein-emergence}). On the other hand, the
convergence point may remain stable for a while---i.e.  observable in
experiments---and, we predict, the strand connecting it to the midvein
may emerge according to slightly different
dynamics. \reffig[A]{fig:margin-to-midvein} shows that the largest
$\Delta c$ is near the midvein while the peak of concentration is at
the convergence point. Accordingly, the new strand will emerge at the
midvein first and make its way to the convergence point---just like a
new strand in an areole progresses from the existing vein toward the
peak of concentration (e.g., see \cite{pdswz:paper}).

\begin{figure}
  \begin{center}
    \begin{tabular}{c}
      \includegraphics{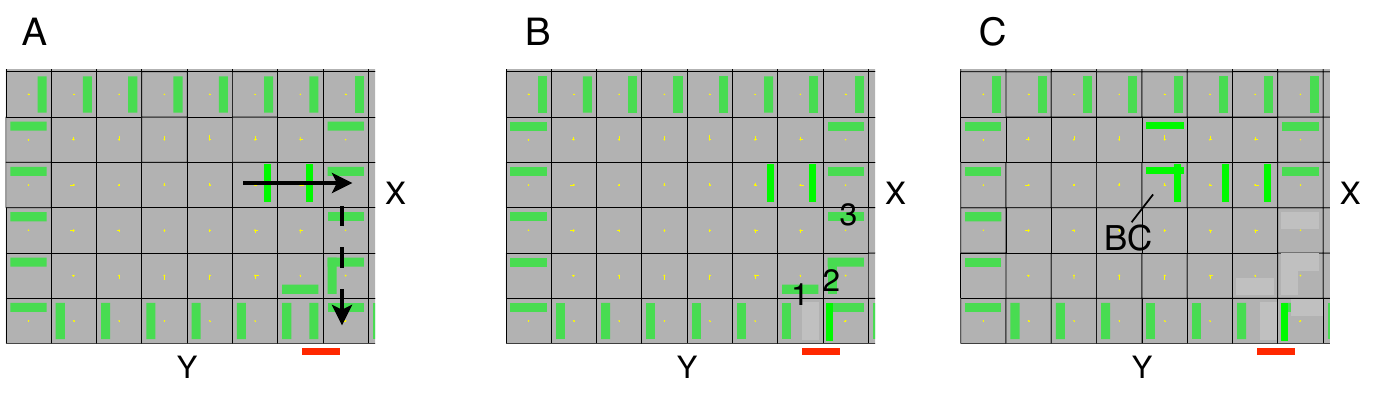}
    \end{tabular}
  \end{center}
   \caption[Formation of first loop.]{\label{fig:final-loop} Formation
     of first loop. \FP{A} The new PIN expression as predicted by our
     model is shown (solid arrow). This increases auxin supply into
     the strand along $X$, which causes the concentration to increase
     there. Since the midvein cannot drain the new infusion of auxin
     immediately---adaptation time---there is a back propagation
     effect (dashed arrow) which causes the concentration of auxin to
     increase even at the convergence point (CP). \FP{B} The bipolar
     cell disappears since the PIN between BC and CP flips sides due
     to the increased $\Delta c$ caused by the new strand. As this
     happens, the PIN at interface 1 starts pushing auxin against the
     gradient and eventually disappears. Next, interfaces 2 and 3
     disappear for the same reason. \FP{C} The new strand meets in the
     middle of the areole---where the concentration peak is---creating
     a bipolar cell there (BC) but the connection to the margin is now
     lost.}
\end{figure}

In fact, dynamics much like the ones discussed for areoles predict the
next step in the elaboration of the first loop. The PIN expression
domain in \reffig[B]{fig:margin-to-midvein} that includes the portion
of the margin between the tip and the convergence point, the midvein
and the newly created strand form, geometrically, an
areole. Surprisingly, the distribution of auxin concentration in this
geometric areole looks very much like the distributions from
\cite{pdswz:paper} and \cite{plos-paper1} as
\reffig[C,D]{fig:margin-to-midvein} illustrate: low concentration
along boundary, peak inside, and highest $\Delta c$ at the
boundary. The new strands inside this geometric areole can be
predicted by following the largest $\Delta c$ as discussed in
\cite{pdswz:paper} and the companion paper \cite{plos-paper1}---we
obtain \reffig{fig:final-loop}.

The key difference, however, is that the boundary PED of the areole is
not entirely stable. The interface of the bipolar cell with the
convergence point actually expresses PIN in the direction against
diffusion. Thus, a sufficiently quick infusion of auxin into the
convergence point cell can increase $\Delta c$ enough to cause a flip,
remove the bipolar cell, and reestablish the continuity of PIN1
polarity along the margin. Such an infusion can result from the
creation of the new strand.

Indeed, as \reffig{fig:final-loop} illustrates, the extension of the
PED in the geometric areole can remove PIN from some interfaces. As
the new PIN expression emerges (\reffig[A]{fig:final-loop}), the right
hand side strand is infused with additional auxin. Even though the
strand may potentially adapt to the new demands---by increasing the
amounts of PIN that drain the hormone away---any adaptation requires
time during which the whole strand will see higher overall
concentrations. This is especially important at the convergence point
where, in particular, the $\Delta c$ through the interface with the
bipolar cell (red bar in \reffig{fig:final-loop}) will rise. Since, as
we argued in \cite{plos-paper1}, the new strand (solid arrow) is
likely to extend in quick bursts, the infusion of auxin into the CP is
likely to increase this $\Delta c$ and flip the PIN expression. Once
that happens, the PED connecting the margin to the midvein will
gradually disappear as well---the stability of the convergence point
requires all of the supporting structure as illustrated in
\reffig{fig:bipolar-formation}.

\section{PIN Patterning in the {\em Arabidopsis} Embryo}
\label{sec:embryo}

\begin{figure}
  \centering
  \includegraphics{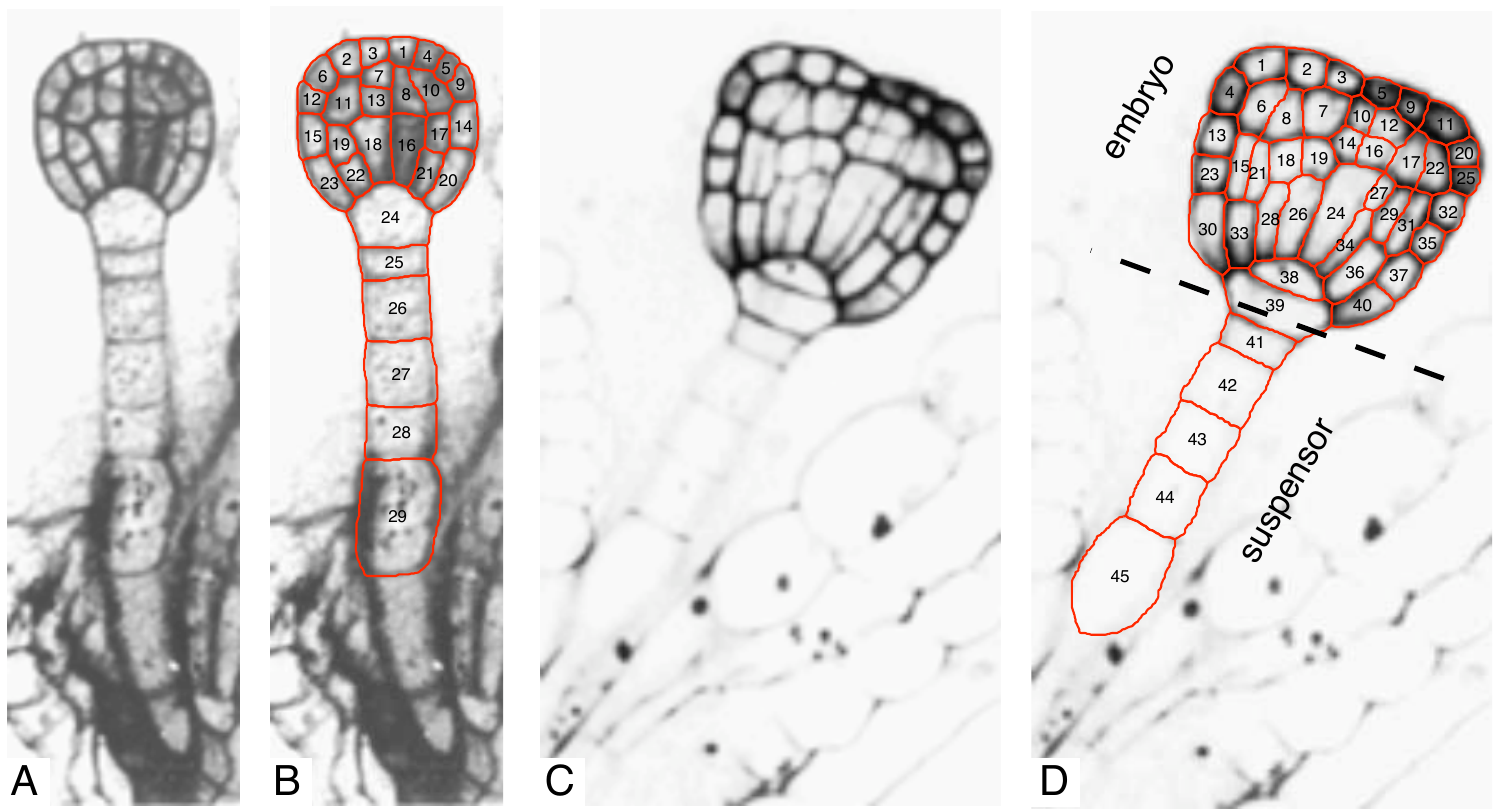}
  \caption[Obtaining the geometry of {\em Arabidopsis}
    embryos.]{\label{fig:embryo-tracefig} Obtaining the geometry of
    {\em Arabidopsis} embryos. \FP{A} Globular stage (from
    \cite{berleth-chatfield-2002}). \FP{B} Domain extraction from
    \CP{A} following the procedure in
    \refsec{sec:voronoi-explanation}. Embryo: cells 1 through 23;
    suspensor: cells 24 through 29. \FP{C} Triangular stage (from
    \cite{berleth-chatfield-2002}).}
\end{figure}

\begin{figure}
  \centering
  \includegraphics[height=8in]{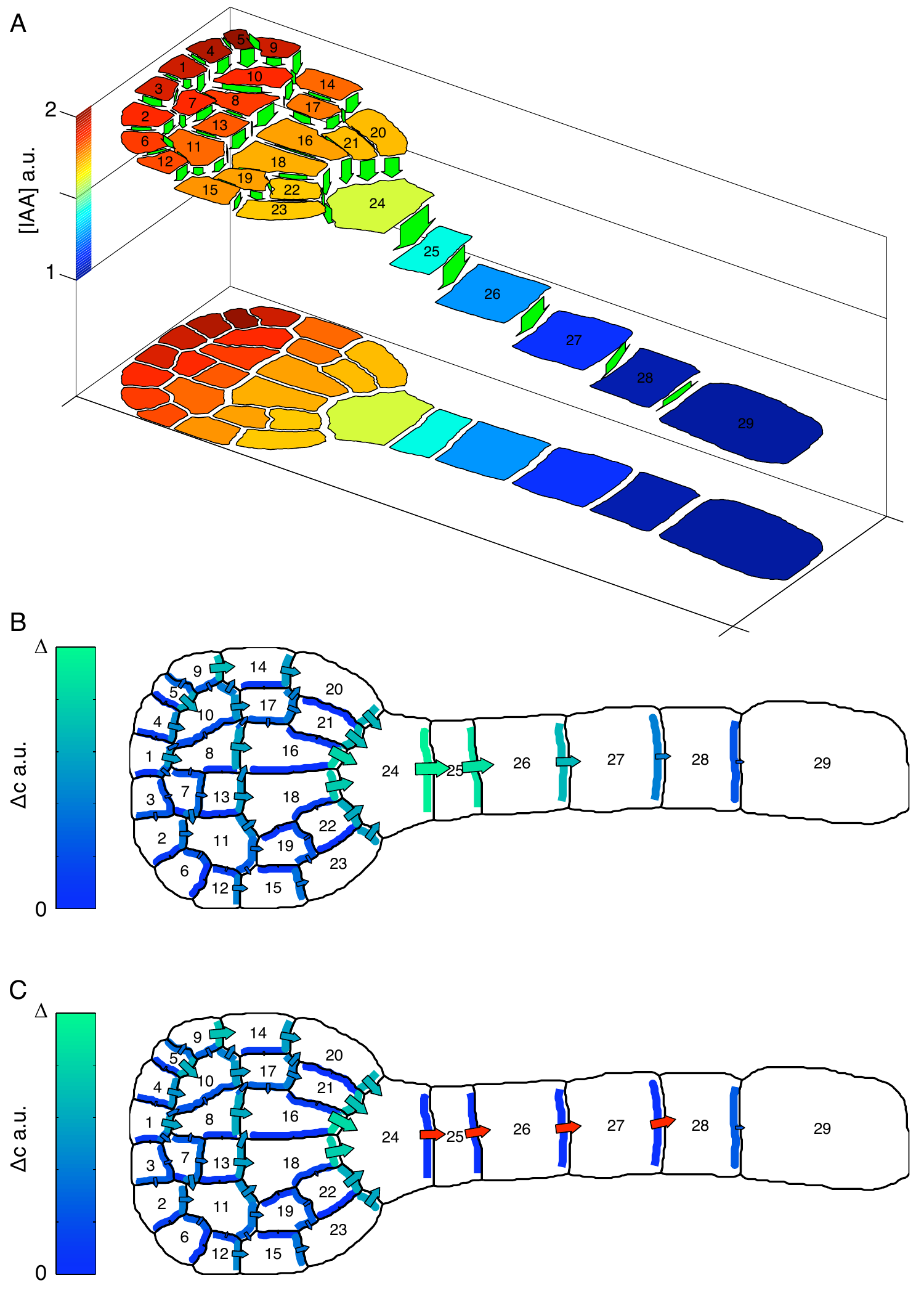}
  \caption[Prediction of PIN pattern formation in {\em Arabidopsis}
    embryos.]{\label{fig:embryo-steps01} Prediction of PIN pattern in
    embryos. \FP{A} Concentration of auxin assuming Helmholtz
    conditions, i.e. no polar transport. \FP{B} Differences of
    concentration, $\Delta c$, for auxin distribution as in
    \CP{A}. Arrow size proportional to $\Delta c$. Note largest
    $\Delta c$ between suspensor cells 24 and 25: PIN should appear
    there first. \FP{C} $\Delta c$ after introducing some of predicted
    polar transport (red arrows). Note large $\Delta c$ between
    suspensor cell 24 and the embryo: PIN should form there next.}
\end{figure}

\begin{figure}
  \centering
  \includegraphics{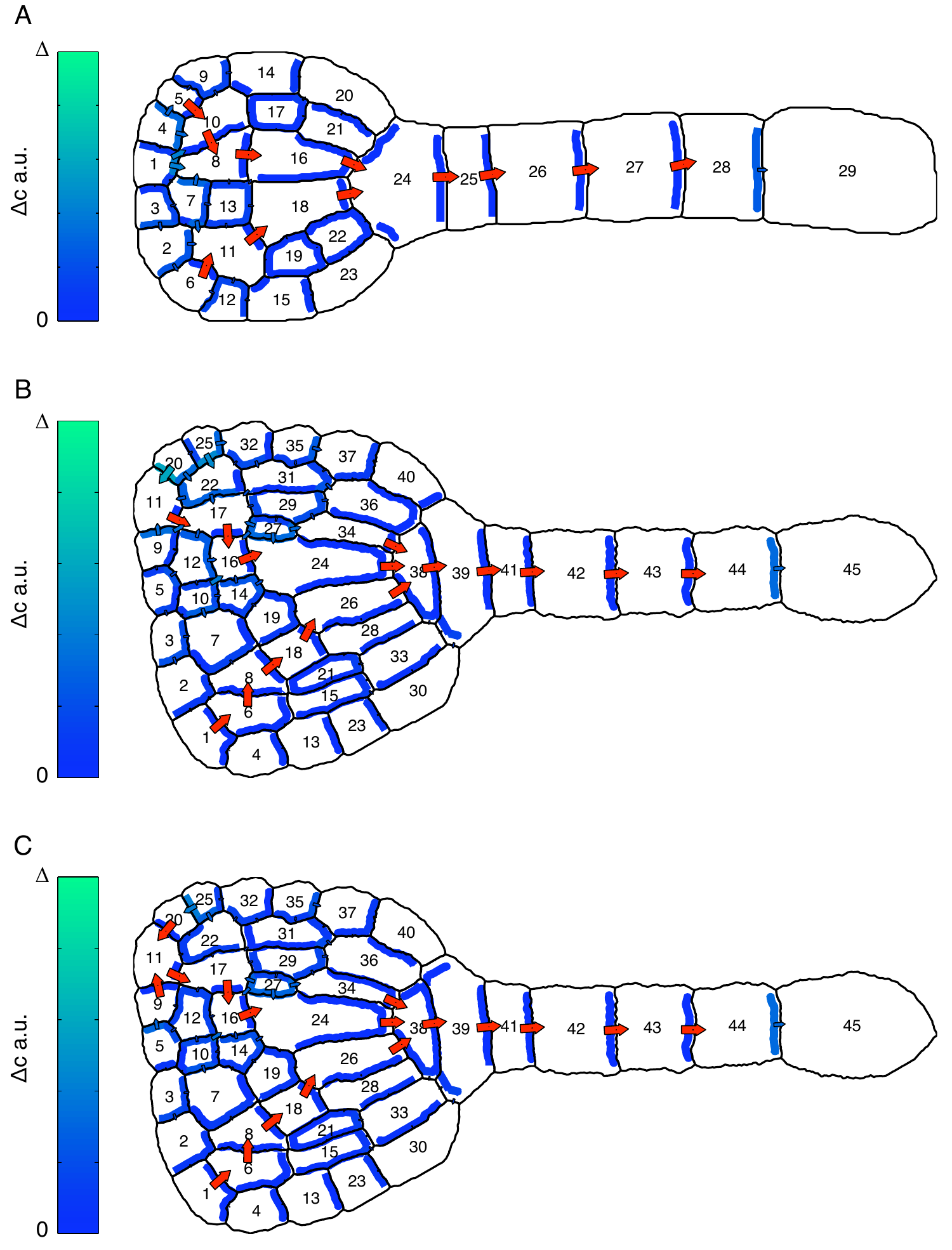}
  \caption[Prediction of PIN pattern formation in {\em Arabidopsis}
    embryos (continued).]{\label{fig:embryo-steps234} Prediction of
    PIN pattern formation in embryos (continued). \FP{A} Predicted PIN
    pattern by following $\Delta c$. No more large $\Delta c$
    exist. \FP{B} $\Delta c$ in a triangular stage embryo with
    established PIN pattern. Note that the new cells (size
    distributions) create a non-trivial $\Delta c$ between cells 20
    and 11: PIN should form there next. \FP{C} Final PIN pattern
    predicted by our model. Note the `convergence point' at cell
    11. Blue (color coded) arrows show $\Delta c$---size represents
    value. Red arrows represent polar transport---their size does not
    indicate strength.}
\end{figure}

The same Non-linear Polar Transport model which accounts for many key
events during vein formation also explains key patterning events at
the earliest stages of plant development: the embryo
formation. \reffig{fig:embryo-tracefig} shows how to obtain the
geometry of embryos from images of the tissues. Then, in
\reffig{fig:embryo-steps01} and \reffig{fig:embryo-steps234} we show
how the polar pattern of PIN expression is predicted by our model.

We begin with the assumption that no polar pattern exists. Uniform
expression of PIN is equivalent to diffusion transport---as we argue
in Appendix~\refsec{sec:chemiosmotic-active-transport}---so our Helmholtz
model suffices to predict the initial distribution of auxin
concentration for an embryo in the globular stage,
\reffig[A,B]{fig:embryo-steps01}. This distribution exhibits a maximal
difference in concentration between the suspensor cells numbered 24
and 25 which, according to Schema 1, makes that interface the most
likely first location for strong PIN expression. As this happens, the
$\Delta c$ there decreases and causes PIN to appear in several other
suspensor interfaces, as shown in \reffig[C]{fig:embryo-steps01}. In
turn, this event causes the $\Delta c$ between the suspensor cell 24
and its neighboring embryo cells to increase significantly, and
polarizes the expression of PIN at those interfaces. This polarization
continues into the embryo---like the formation of a new strand in an
areole---until it reaches the smallest cells: cell 5 and cell 6 in
\reffig[A]{fig:embryo-steps234}. At that point no significant $\Delta
c$ exists in the globular stage embryo.  The triangular stage,
resulting from several cell divisions, presents a different cell size
distribution and significant $\Delta c$ appear again,
\reffig[B]{fig:embryo-steps234}. In fact, a convergence point appears
near cell 11 due to the established pattern of PIN from the earlier
stages.

The final predicted pattern, shown in \reffig[C]{fig:embryo-steps234}
and \reffig[C]{fig:embryo-summary}, captures two important qualitative
characteristics observed in the lab: one regarding the emergence of
new leaves and one regarding the root tip. The theory developed here
explains the initial formation of the `convergence point' patterns
that mark the appearance of new leaf primordia. In the case of the
embryo, these are the cotyledons, but the same process could be taking
place near the shoot meristem in mature plants. The theory also
explains why the polarity of PIN in the root is towards the root
tip. That is the result of the original differences of cell size. The
suspensor cells, being larger, produce auxin at a lower per-volume
rate than the embryo cells, so suspensor cells are predicted to have a
lower concentration of auxin. Thus, the difference in concentration
sets up the polar pattern near the root tip, which will be maintained
through development due to the memory-like behavior of PIN
dynamics. In a companion paper \cite{plos-paper3} we shall show that
this initial pattern sets the stage for the observed `reflux' pattern
in mature roots, and that the Polar Transport model developed here
explains the events that lead to the mature pattern as well.

\begin{figure}
  \centering
  \includegraphics{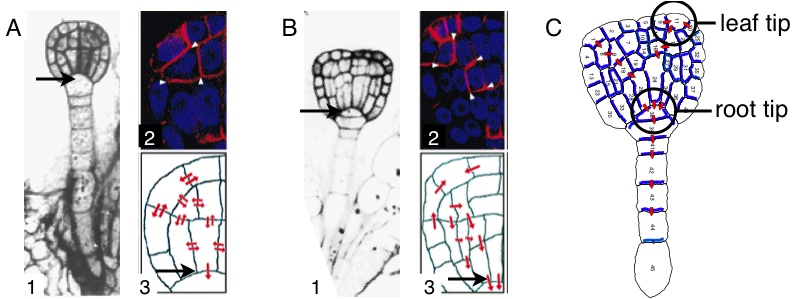}
  \caption[Prediction of polar expression of PIN in {\em Arabidopsis}
    embryos and comparison to real
    measurements.]{\label{fig:embryo-summary} Prediction of polar
    expression of PIN in {\em Arabidopsis} embryos and comparison to
    real measurements. \FP{A} Globular stage. \FP{B} Triangular
    stage. Note that the suspensor cells are larger than the embryo
    cells. Hence, our Helmholtz model predicts higher concentration in
    the embryo than in the suspensor. Black arrows point at where the
    root tip will be. Insets 1 from \cite{berleth-chatfield-2002};
    insets 2 and 3 from \cite{benkova-auxingradients}. \FP{C} Our
    model predicts the typical polar transport patterns in both the
    leaf primordium and near the root tip: (1) the `convergence point'
    pattern near the embryo tip where a leaf primordium will emerge (a
    cotyledon); and (2) the polarity of PIN expression is toward the
    root tip and due to the larger suspensor cells. }
\end{figure}

\section{Conclusion}


The diversity of phenomenology, data, and temporal information about
plant development is difficult to unify unless an abstract perspective
is taken. In this series of two papers we sought to articulate several
of the basic principles that could underlie the developmental biology
of plants, and to place these within a mathematical framework that
allowed their consequences and implications to be calculated.

Realizing that distance information is fundamental, we introduced
three principles that could work together within a reaction-diffusion
model to predict how global distance information could be signalled
locally, at the cell level.  The first two principles, that auxin is
produced at a constant rate in each cell and is destroyed in
proportion to concentration, yielded models that predicted qualitative
auxin distributions. Together with a simple schematic rule that cells
begin to convert from a ground to a vascular state provided the
concentration difference exceeds a limiting value led to a model of
vascular formation.

In this paper we elaborated this schema to a more realistic
biophysical level.  Facilitated transport of auxin by carriers such as
those in the PIN and AUX families serves to effect the
ground-to-vascular transition, but it does so in the context of
non-linear dynamics with substantial predictive power. While it is
widely accepted that PIN patterns prefigure vascular patterns, we were
able to show that this dynamic--- including the discrete and
specialized structures that arise within it during plant
development---can be predicted. Simulations showed how the leaf
primordium emerges, how convergence points arise on the leaf margin,
how the first loop is formed, how the intricate pattern of PIN shifts
during the early establishment of vein patterns in incipient leaves of
Arabidopsis, and how the midvein forms. Most importantly, we showed
how the embrio provides an initial configuration sufficient to get it
all started.

Just as principles provide the foundation for theoretical physics, and
mathematics provides the machinery to realize their consequences, we
have sought to provide an abstract view of the development of pattern
structure in plants. Our principles were designed to articulate key
aspects of plant biology when viewed from an abstract, ``high in the
air'' perspective.  We believe they will hold as the facts relating to
auxin production, destruction, and facilitated transport develop
further; and that their predictions will lead to further experiments
revealing the beauty and subtlety of plant development.

\appendix

\section{Simplification of the Chemiosmotic Model}
\label{sec:chemiosmotic-active-transport}

\begin{figure}[t]
  \begin{center}
    \includegraphics{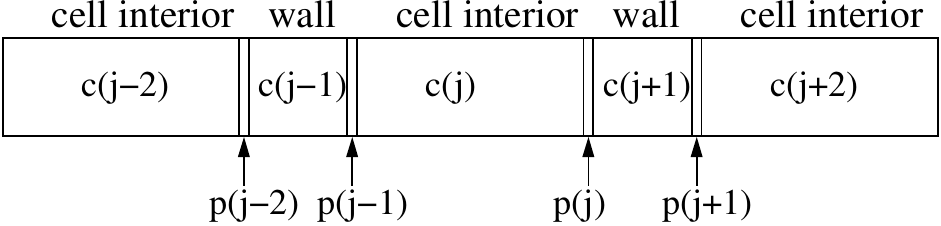} \\
  \end{center}

  \caption[Schematic depiction of
    cells.]{\label{fig:cell-int-wall-setup} Schematic depiction of
    cells (compare with \cite[Fig. 1]{goldsmith-1977}). Five major
    compartments are shown: three cell interiors and two cell
    walls. The permeabilities through the plasmalemma are shown for
    all four compartment junctions and denoted by $p(i)$.}
\end{figure}

We now show that the net effect of active transport as predicted by
the chemiosmotic theory may be captured by a small extension to our
earlier model. Specifically, we show that the same effect may be
obtained without the need to introduce separate compartments for cell
interiors and cell walls---it suffices to keep the interior of cells
as discrete compartments and to introduce the effect of carriers as
directional active transport coefficients. These coefficients are a
direct measure of the net effect of facilitated auxin transport, which
we develop in the main body of the paper.


Suppose, then, that we have a system of five compartments as in
\reffig{fig:cell-int-wall-setup}. Let $J(j\to j+1)$ denote the mass
flux between compartments $j$ and $j+1$, so $J(j\to j+1)=-J(j+1\to
j)$. Define $\Phi(j\to j+2)$ as
\begin{equation}
\Phi(j\to j+2) = \frac{J(j\to j+1) + J(j+1\to j+2)}{2} .
\label{eq:Phi}
\end{equation}

\noindent
Our task is to show that it is possible to think of $\Phi(j\to j+2)$
only in terms of the compartments $j$ and $j+2$, the cell interiors.

We do this by examining the behavior of the system at equilibrium. The
substance is only produced inside cells and it is also depleted there
by various metabolic processes. The following equations formalize
these assumptions:
\begin{equation}
  \begin{array}{llcl}
    (1) & J(j-2\to j-1) + J(j\to j-1) &=& 0 \\
    (2) & J(j-1\to j) + J(j+1\to j) + \rho(j) - \alpha c(j) &=& 0 \\
    (3) & J(j\to j+1) + J(j+2\to j+1) &=& 0 \\
    (4) & J(j-1\to j-2) + \rho(j-2) - \alpha c(j-2) &=& 0 \\
    (5) & J(j+1\to j+2) + \rho(j+2) - \alpha c(j+2) &=& 0 
  \end{array}
  \label{eq:CT}
\end{equation}

\noindent
The concentration of the substance in each compartment is assumed to
be uniform and is denoted by, e.g., $c(j)$ for the middle cell
interior. Each equation corresponds to a compartment. The first three
describe the cell wall $c(j-1)$, the middle cell
interior $c(j)$, and the second cell wall $c(j+1)$. The last two
equations describe the equilibrium dynamics of the leftmost and
rightmost cell interiors, respectively. The production of the
substance is captured by $\rho$ and the destruction by 
$\alpha c$ with $\alpha$ a constant. Let $T(j) = \rho(j) -\alpha c(j)$
and write the dynamics (at equilibrium) of the cell interiors assuming
that the cell membranes and cell walls together are a single interface
and that $\Phi$ describes the flux through those interfaces:
\begin{equation}
  \begin{array}{lllcl}
    (a) & \Phi(j-2 \to j) + \Phi(j+2\to j) +& T(j) &=& 0 \\
    (b) & \Phi(j \to j+2) + &T(j+2) &=& 0 \\
    (c) & \Phi(j \to j-2) + &T(j-2) &=& 0 \\
  \end{array}
  \label{eq:CTPhi}
\end{equation}

\noindent
It is easy to check that \refeq{eq:CTPhi} can be obtained from
\refeq{eq:CT} by the following linear combinations:
$$
  \begin{array}{rcl}
    (a) & := & \frac{(1) + (2) + (2) + (3)}{2} \\
    (b) & := & \frac{(3) + (5) + (5)}{2} \\
    (c) & := & \frac{(1) + (4) + (4)}{2}
  \end{array}
$$

\noindent
Thus, if the $\Phi$ flows only depend on the concentrations in three
of the five compartments, then \refeq{eq:CTPhi} has a unique solution
which is compatible with the chemiosmotic theory. We now derive the
conditions under which this holds and, therefore, demonstrate that our
simulations are in agreement with the chemiosmotic theory.


The substance is a weak acid and that the concentration $c = [HA] +
[A^-]$, i.e. the total concentration of protonated (associated) acid
and the concentration in ionic form (dissociated). The dynamics of $c$
depend on the dynamics of both $[HA]$ and $[A^-]$, so we need to
express each one as a function of $c$. Noting that the dissociation of
a weak acid is given by the reaction
$$
HA \rightleftharpoons H^+ + A^- ;\quad
K_a = \frac{ [H^+] [A^-] }{ [HA] } ~\mathrm{mol}\cdot \mathrm{l}^{-1}
$$
\noindent
we write
$$
K = \frac{[H^+] [A^-]}{[HA]} \implies
[HA] = \frac{10^{-pH} [A^-]}{10^{-pK}} \implies
[A^-] = [HA] 10^{pH-pK}
$$
which together with $c = [HA] + [A^-]$ gives
\begin{equation}
  [HA] = \frac{1}{10^{pH-pK} +1}c \quad \mathrm{and} \quad
  [A^-] = \frac{10^{pH-pK}}{10^{pH-pK} +1}c
  \quad .
  \label{eq:ha-a}
\end{equation}

\noindent
The change of total concentration due to substance transport is the
result of two simultaneous effects: (1) the protonated form $HA$
diffuses through the membrane, and (2) the anion $A^-$ moves across
the membrane as a result of auxin carriers and the electrical
potential established by the difference in $pH$ between the interior
and exterior (cell wall) of the cell \cite{goldsmith-1977}.  This
dynamical behavior is therefore (see, e.g.,
\cite{goldsmith-etal-1981}):
\begin{equation}
  J(i\to e) = P_{HA} \paren{[HA]_e - [HA]_i} + P_{A^-}g(V)
  \paren{[A^-]_e - [A^-]_i f(V)} 
\label{eq:ct-eq}
\end{equation}

\noindent
where $P_{HA}$ is the permeability of the protonated acid; $P_{A^-}$
is related to the amount of auxin carriers on the membrane; $V$ is the
membrane voltage; the subscript $e=j+1$ and $i=j$ denote the exterior
and the interior of a cell, resp.; $f(V) = \exp(\phi)$ for
$\phi=-FV/RT$, an electrical term consisting of the Faraday constant
$F$, the gas constant $R$, and absolute temperature $T$; $g(V) =
\frac{\phi}{1-f(V)}$. We do not include the diffusion of $A^-$ through
the membrane because it appears to be negligible \cite{bean-etal-1968,gutknecht-tosteson-1973,gutknecht-walter-1980}.

Let $\beta_j = 1/(10^{pH_j-pK}+1)$ and $\alpha_j=1-\beta_j$.  Let
$h=g(V) f(V)$ and, using \refeq{eq:ha-a}, rewrite \refeq{eq:ct-eq} as
$$ 
J(j\to j+1) = P_{HA} \paren{\beta_j c(j) - \beta_{j+1}c(j+1) } +
  P_A(j) g \paren{ \alpha_j c(j) - f \alpha_{j+1} c(j+1)  }
$$

It is reasonable to assume that $f_j = f_{j+2}$ because both functions
relate the interior to the exterior of the cell in the same order and
the voltage appears to be the same. However, the voltage has opposite
sign when the order is reversed, so that $f_j = -f_{j+1} =
1/f_{j+1}$. Therefore, setting $f=f_j$, we obtain
$$ 
J(j+1\to j+2) = P_{HA} \paren{\beta_{j+1} c(j+1) - \beta_{j+2}c(j+2) } +
  P_A(j+1) f g \paren{ \alpha_{j+1} c(j+1) - \frac{1}{f} \alpha_{j+2} c(j+2)  }
$$

We can now write $2\Phi(j\to j+2) = J(j\to j+1)+J(j+1\to j+2)$ as
\begin{equation}
\begin{array}{rcl}
2\Phi(j\to j+2) = P_{HA} \paren{ \beta_{j}c(j)-\beta_{j+2}c(j+2)  }
&+& P_A(j) g \alpha_j c(j) + \paren{ P_A(j+1)-P_A(j)} fg
\alpha_{j+1}c(j+1) \\
&-& P_A(j+1) g \alpha_{j+2} c(j+2)
\end{array}
\label{eq:mess1}
\end{equation}

Substituting $P_A(j) = P_A + p(j)$ we can rewrite \refeq{eq:mess1} as
\begin{equation}
  \begin{array}{rcl}
    2\Phi(j\to j+2) = && P_{HA} \paren{ \beta_{j}c(j)-\beta_{j+2}c(j+2)
    } +
    P_A g \paren{ \alpha_j c(j) -  \alpha_{j+2} c(j+2) } \\
    &+& p(j) g \alpha_j c(j) - p(j+1) g \alpha_{j+2} c(j+2) 
    + \paren{ p(j+1)-p(j)} f g  \alpha_{j+1} c(j+1)
  \end{array}
  \label{eq:mess2}
\end{equation}

Now if $p(j+1)=p(j)$---in case both sides of the cell have the same
permeabilities---then the last term becomes zero. In fact, even if
the permeabilities are different the term can be ignored because $fg
\sim \frac{1}{100}$ according to the best available estimates
(i.e. $V=-120$mV). Also, since the $pH$ difference between the inside
and the outside of a cell is assumed to be the same for all cells
(metabolically maintained), we can write $\beta_j = \beta_{j+2}=\beta$
and $\alpha = 1-\beta$. \refeq{eq:mess2} becomes
$$ 2\Phi(j\to j+2) \sim P_{HA} (1-\beta) (c(j)-c(j+2)) + P_A g \beta
(c(j)-c(j+2)) + g \beta ( p(j) c(j) - p(j+1) c(j+2))
$$
\noindent
so assuming w.l.o.g. that $p(j+1)=0$ (i.e. because $P_A=P(j+1)$) and
collecting terms
\begin{equation}
2\Phi(j\to j+2) \sim (c(j)-c(j+2)) \underbrace{\paren{ P_{HA}
    (1-\beta) + P_A g \beta} }_\mathrm{Diffusion~coeff.} +
\underbrace{g \beta p(j) c(j)}_\mathrm{Polar~transport}
\end{equation}

Hence, the flux of auxin between cells can be written as the flux
between cell interiors as
\begin{equation}
2\Phi(j\to j+2) \sim D (c(j)-c(j+2)) + P(j) c(j)
\label{eq:finalsimp}
\end{equation}

\noindent
Note that if the auxin carriers are homogeneously distributed on the
cell membrane or if there are none, then $P(i)=0$ and the flux is
governed by Fick's Law. If, however, the carriers are only located on
one side of the cell, then $P_A=0$ and $P(i)$ in \refeq{eq:finalsimp}
describes the effect completely. In particular, $P(i)$ is proportional
to the amount of carriers on the portion of membrane of cell $j$
adjacent to cell $j+2$ (as in \reffig{fig:cell-int-wall-setup}).

\section{Midvein PED prediction}

Here is the algorithm used in \refsec{sec:midvein-emergence}.  It is
possible to predict the length of the midvein by measuring the length
(or size) of the marginal PIN1 expression, the MPEDs. We fix a
threshold $\tau_1$ on $\Delta c$ for the formation of new PIN1 and
measure the MPEDs. We obtain $c$ as in \refsec{sec:non-linear-model}
by first assuming that PIN1 works against diffusion in MPEDs---mode
1---and with diffusion in the midvein PED (if such a PED
exists)---mode 2. We check each interface where PIN1 exists against
the computed $c$ and change its operation mode accordingly. If all
interfaces are consistent with $c$, then we find the largest $\Delta
c$ in the domain and check whether it is above $\tau_1$. We extend the
PED if $\Delta c>\tau_1$; otherwise, we finish the procedure. This
procedure is collected in \refalg{alg:midvein-from-mped} and
illustrated in \reffig{fig:pin-veins-midvein-1}. We can predict the
new PEDs in this way provided that the marginal expression of PIN1 is
known.

\begin{algorithm}
  \algsetup{indent=2em}
  \begin{algorithmic}[1]
    \STATE {\sc input:} Domain with MPED, $\tau_1$
    \STATE {\sc output:} Domain with MPED and midvein PED
    \STATE midveinincreased := true
    \WHILE{midveinincreased == true}
    \STATE midveinincreased := false
    \STATE consistent := false
    \WHILE{consistent == false}
      \STATE consistent := true
      \STATE Compute $c$ as in \refsec{sec:non-linear-model}
      \FORALL{Neighbors $i$ and $j$}
        \IF{$c(i)>c(j)$ and $Mode(i,j)\neq 0$ and $Mode(i,j)\neq 1$}
          \STATE $Mode(i,j) = 1$, consistent:=false
        \ENDIF
        \IF{$c(i)<c(j)$ and $Mode(i,j)\neq 0$ and $Mode(i,j)\neq 2$}
          \STATE $Mode(i,j) = 2$, consistent:=false
        \ENDIF
      \ENDFOR
    \ENDWHILE
    \STATE Find $i,j$ neighbors and $\Delta c=c(i)-c(j)$ where
      $\abs{\Delta c}$ is largest
    \IF{$\Delta c \geq \tau_1$}
      \STATE $Mode(i,j)=1$, midveinincreased=\TRUE
    \ENDIF
    \IF{$\Delta c \leq \tau_1$}
      \STATE $Mode(j,i)=1$, midveinincreased=\TRUE
    \ENDIF
    \ENDWHILE
  \end{algorithmic}
  \caption{\label{alg:midvein-from-mped} Prediction of midvein from
    MPEDs. $Mode(i,j)$ denotes the action of PIN1 from cell $i$ to
    cell $j$ and can be: 0 for no action, 1 if against gradient, and 2
    if with gradient.}
\end{algorithm}

\section{Geometric Domain Definition using Voronoi Diagrams}
\label{sec:voronoi}
\label{sec:voronoi-explanation}

Assuming that auxin diffuses much faster inside the cell than through
the cell wall, we do not need to model auxin transport inside the
cell, only between neighboring cells. This is the basis of the
simulations in Ref.~\cite{plos-paper1}, for example, but the geometry
and topology of real cells is not accurately captured there. In
particular, the neighbor relations are inaccurate as are the sizes of
the interfaces. Our approach based on Voronoi diagrams solves this
problem. We now describe how arbitrary geometric domains with the
required properties can be defined from the Voronoi diagram of a
collection of points, and then demonstrate how images of plant tissues
can be segmented into cells using this approach. The pseudo-code for
the procedure is presented in
Algorithm~\ref{alg:voronoi-cell-trace}.\footnote{The algorithms as
  presented here are fairly inefficient---they run in $O(n^2)$ time
  and space. In practice, $O(n \log n)$ times were achieved with
  appropriate data structures.}

\begin{figure}
  \begin{center}
    
    \includegraphics[]{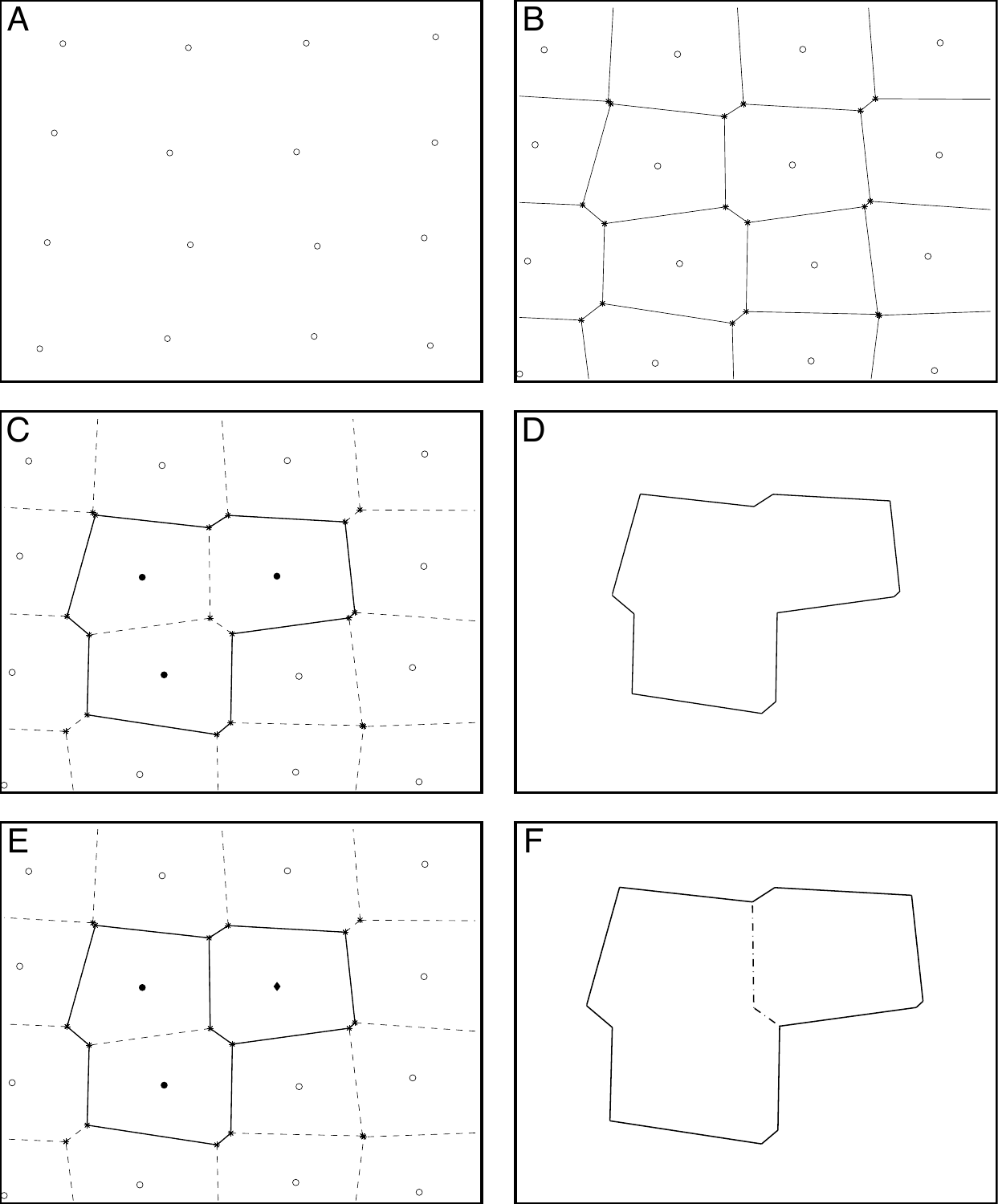}
      \caption[How to use Voronoi diagrams to draw.]{How to use
        Voronoi diagrams to draw. \FP{A} Points in 2-D. \FP{B} Voronoi
        diagram of points in \CP{A}. Voronoi vertices (asterisks) are
        joined by Voronoi edges or segments (lines). \FP{C} Two groups
        of points: open circles and closed circles. Boundary between
        two groups (solid line) represents shape in \CP{D}. \FP{D}
        Shape from \CP{C}. \FP{E} Three groups of points. \FP{F} Shape
        now consists of two regions that share an interface. The size
        of each region can be computed as the sum of sizes for each
        Voronoi cell, and the size of the interface is the sum of the
        sizes for each segment that comprises it.  The Voronoi diagram
        and the convex hull of each Voronoi cell (volume and boundary)
        can be computed using the publicly available software package
        {\tt qhull} or any of the algorithms in
        O'Rourke~\cite{orourke-1998}.}
    \label{fig:voronoi-explanation}

  \end{center}
\end{figure}

Consider \reffig[A,B]{fig:voronoi-explanation}. Two dimensional points
(open circles) are defined and the Voronoi diagram computed. The
diagram consists of straight line segments that join the so-called
Voronoi vertices (represented by asterisks). Each segment is closest
to two exactly points and runs through part of the line that divides
the plane at halfway between the two points. Some points are
surrounded by these segments completely and thus define a closed
region. This region, known as a Voronoi cell, is always convex and it
is possible to determine its size (area).

Now suppose that the points are separated into two groups: the open
circles and the closed circles, as in
\reffig[C]{fig:voronoi-explanation}. Then only some of the segments
(solid lines) will separate the two groups, while the others will not
(dashed lines). In this example, the points represented by a closed
circle form a group that defines a closed region of space that
consists of the union of the Voronoi cells of each of the points. This
larger region (\reffig[D]{fig:voronoi-explanation}) is no longer
convex, but its size (area) can be computed by adding the sizes of the
Voronoi cells. In addition, the size of the boundary of the region can
be computed by adding the lengths of the segments that comprise
it. Note that this procedure can be used to define any region of
space, i.e. a shape, by using more points.

\begin{algorithm}
  \algsetup{indent=2em}
  \begin{algorithmic}[1]
    \STATE {\sc input:} Groups of points $G_1,\dots,G_k$, of which $G_1$ is the
    bounding group; diffusion coefficient $D$
    \STATE {\sc output:} Geometry of $k-1$ cells $CSegs$, size $S(i)$,
    production function $\rho(i)$,
    interface size $I(i,j)$, diffusion matrix $M(i,j)$
    \STATE \{ $vsegs, vcells$ \} $\leftarrow$ {\sc Voronoi}$(\cup_{i=1}^k G_i)$
    \FOR {$i=2$ to $k$}
       \STATE $S(i) \leftarrow 0$
       \FORALL {$P \in G_i$}
       \STATE $S(i) \leftarrow S(i)+${\sc ConvexHullVolume}($vcells(P)$)
       \ENDFOR
    \STATE CSegs$(i-1) \leftarrow$ {\sc ComputeBoundarySegs}$(G_i,
    vsegs, vcells)$ \{See Algorithm~\ref{alg:boundarysegs}\}
    \ENDFOR
    \FOR {$i=1$ to $k-1$}
      \FOR {$j=1$ to $k-1$ and $j\neq i$}
      \STATE $I(i,j) \leftarrow$ {\sc
        ComputeIntersectionLength}$(G_i, G_j, vcells)$ \{ See
      Algorithm~\ref{alg:intersectionlength}\} 
      \STATE $M(i,j) \leftarrow D* I(i,j)$
      \ENDFOR
    \ENDFOR
    \FOR {$i=1$ to $k-1$}
      \STATE $M(i,i) \leftarrow \sum_{j\neq i} - M(i,j)$
      \STATE $\rho(i) \leftarrow 1/S(i)$
    \ENDFOR    
  \end{algorithmic}
  \caption{\label{alg:voronoi-cell-trace} {\sc ComputeSetup}. The
    calculation of the Voronoi diagram as well as of the volume of the
  convex hull of a set of points is documented in
  O'Rourke~\cite{orourke-1998}. The freely available software package
  {\tt qhull} contains efficient implementations for both.}
\end{algorithm}

\begin{algorithm}
  \algsetup{indent=2em}
  \begin{algorithmic}[1]
    \STATE {\sc input:} A group of points $G$, Voronoi segments
    $vsegs$, Voronoi cells $vcells$
    \STATE {\sc output:} Segments $segs$ that form the boundary of $G$
    with other groups.
    \STATE $segs \leftarrow \emptyset$
    \FOR {$seg\in vsegs$}
      \IF {$P\in G, Q\not\in G$ s.t. $seg\in vcell(P)$ and $seg\in vcell(Q)$ }
      \STATE $segs \leftarrow segs \cup \{seg\}$
      \ENDIF
    \ENDFOR
  \end{algorithmic}
  \caption{\label{alg:boundarysegs} {\sc
      ComputeBoundarySegs}$(G, vsegs, vcells)$.}
\end{algorithm}

\begin{algorithm}
  \algsetup{indent=2em}
  \begin{algorithmic}[1]
    \STATE {\sc input:} A group of points $G$, Voronoi segments
    $vsegs$, Voronoi cells $vcells$
    \STATE {\sc output:} Length of intersection $len$.
    \STATE $len \leftarrow 0$
    \FOR {$seg\in vsegs$}
      \IF {$P\in G_i, Q\in G_j$ s.t. $seg\in vcell(P)$ and $seg\in vcell(Q)$ }
      \STATE $len \leftarrow len+${\sc Length}$(seg)$
      \ENDIF
    \ENDFOR
  \end{algorithmic}
  \caption{\label{alg:intersectionlength} {\sc
      ComputeIntersectionLength}$(G_i, G_j, vcells)$.}
\end{algorithm}

Now suppose that there are three groups of points as in
\reffig[E]{fig:voronoi-explanation}: open circles, closed circles, and
closed diamonds. This configuration defines two closed regions of
space (\reffig[F]{fig:voronoi-explanation}) which share some of the
Voronoi segments (dashed line). As before, the size of each region can
be determined and the size of the interface between the two region can
be computed as well. Thus, this method can not only define any shape
but it can also provide both the shape size (area) and the size of the
interface between any two neighboring shapes. These were the
requirements for a collection of plant cells.

\begin{figure}
  \begin{center}
    
    \includegraphics[]{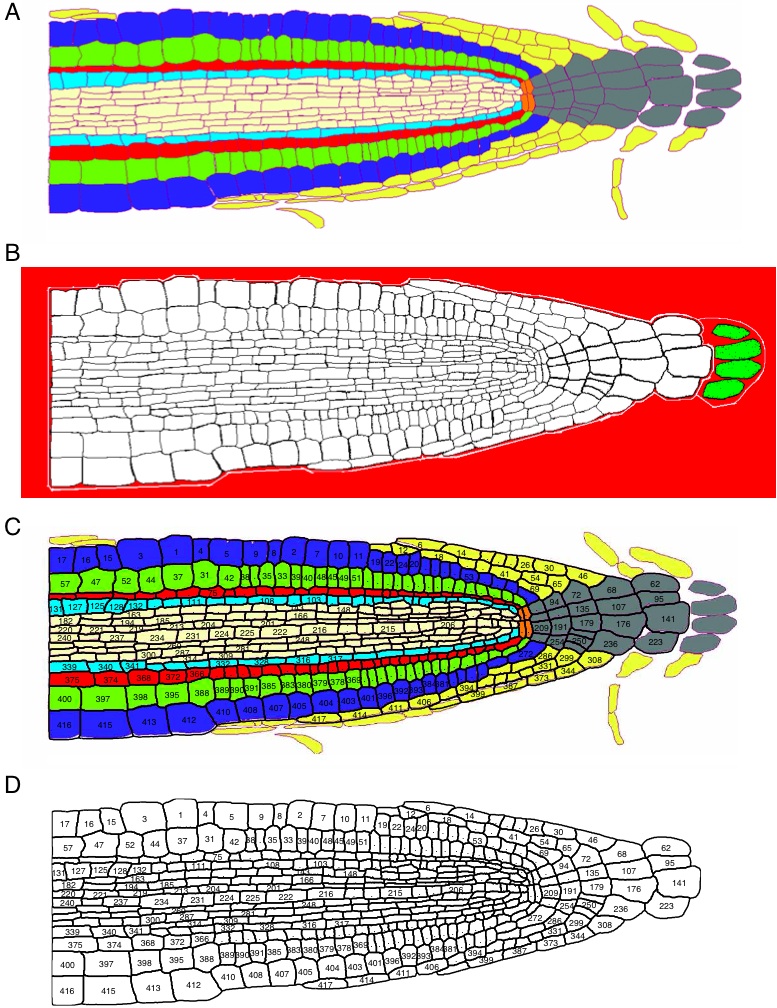}
      \caption[Extracting the geometry and topology of a traced the
        root slice.]{Extracting the geometry and topology of a traced
        the root slice. \FP{A} Image of a hand-traced root slice by an
        expert. \FP{B} Processed image. White pixels in the interior
        of cells, other colors for display purposes and to separated
        different cells. \FP{C} Result of procedure overlaid on top of
        original image. \FP{D} Result of procedure. All cells are
        numbered but a dot ``.'' is displayed for small cells.}
    \label{fig:voronoi-example}

  \end{center}
\end{figure}

\reffig{fig:voronoi-example} shows how this procedure is used to
compute the geometry and topology of a traced root. The original image
is of the root cell structure in a slice of the {\em Arabidopsis} root
hand-traced by an expert. In \reffig[B]{fig:voronoi-example}, we
determine those pixels that belong to the interior of cells and color
them white. Pixels with other colors separate different cells but are
otherwise not used in the computation. We draw a bounding white curve
around the root the pixels of which correspond to the points denoted
by open circles in \reffig{fig:voronoi-explanation}. We extract the
coordinates of each white pixel and group the resulting points by
computing the contiguous regions of white pixels. Thus, each cell
interior is represented by a different (but unique) group of points
and the bounding white curve is the only additional group. Then we
follow the procedure above and obtain the shapes of each cell as well
as their sizes and interfaces. The bounding curve is necessary in
order to make sure that all Voronoi cells inside plant cells are
finite regions of space.

This same procedure works in higher dimensions as well. For example,
the Voronoi diagram of points in three dimensions can be computed
efficiently and three dimensional shapes (representing plant cell) can
be obtained exactly similarly. The notion of cell size then becomes
volume, and interface size becomes area. Both of these can be computed
using the values from Voronoi cells as in the two-dimensional case
discussed above. Thus, this approach produces the required graph
structure for three-dimensional cell measurements as well.

\singlespacing
\bibliographystyle{abbrv}    
\bibliography{paper}

\end{document}

%% file: definitions.tex
\def\FP#1{{\bf ({#1})}}
\def\CP#1{{\em {#1}}}

\newtheorem{thm}{Result}

\newtheorem{prop}[thm]{Proposition}

\newtheorem{hypothesis}{Hypothesis}

\def\implies{\Longrightarrow}

\def\etal{{\em et~al.}}

\def\grad{\ensuremath{\nabla}}
\def\del{\ensuremath{\partial}}

\newlength{\bigfigwidth}
\newlength{\figwidth}
\theoremstyle{definition}

\theoremstyle{remark}

\def\paren#1{\ensuremath{\left( {#1} \right)}}

\def\abs#1{\ensuremath{\left| {#1} \right|}}

\def\del{\ensuremath{\partial}}

\def\grad{\ensuremath{\nabla}}

\newcommand{\reffig}[2][{}]{Fig.~\ref{#2}{\em #1}}

\def\refsec#1{Section~\ref{#1}}
\def\refalg#1{Algorithm~\ref{#1}}

\def\refeq#1{Eq.~\ref{#1}}

\def\c1D{\ensuremath{ {c_{1D}}} }

\def\v#1{\ensuremath{\mathbf{#1}}}

